\documentclass[11pt,a4paper]{article}

\usepackage{authblk}
\usepackage{graphicx}
\usepackage{dcolumn}
\usepackage{bm}

 \usepackage{subfigure}
 \usepackage{physics}
 \usepackage{dsfont}
 \usepackage{eucal}

\usepackage{tabulary}
\usepackage{amsmath}
\usepackage{color} 
\usepackage{float}
\usepackage{amsfonts}
\usepackage{amssymb}
\usepackage{mathtools}
\usepackage{dsfont}
\usepackage{eucal}
\usepackage{braket}
\usepackage{longtable}
\usepackage[left=2cm,right=2cm,top=3cm,bottom=2.5cm]{geometry}
\usepackage{hyperref}
\hypersetup{
    colorlinks,
    linkcolor={black},
    citecolor={black},
    urlcolor={black}
}
\usepackage{caption}
\usepackage{multirow}
\usepackage{enumitem}
\setlist[itemize]{leftmargin=*}

\title{Optimal Control Theory Techniques for Nitrogen Vacancy Ensembles in Single Crystal Diamond}
\author[1,2,*]{Madelaine S.Z. Liddy}
\author[1,3]{Troy Borneman}
\author[1]{Peter Sprenger}
\author[1,4]{David Cory}
\affil[1]{The Institute for Quantum Computing, University of Waterloo, 200 University Avenue, Waterloo, N2L 3G1, ON, Canada}
\affil[2]{Department of Electrical and Computer Engineering, University of Waterloo, 200 University Avenue, Waterloo, N2L 3G1, ON, Canada}
\affil[3]{High Q Technologies, 485 Wes Graham Way, Waterloo, N2L 6R2, ON, Canada}
\affil[4]{Department of Chemistry, University of Waterloo, 200 University Avenue, Waterloo, N2L 3G1, ON, Canada}
\affil[*]{Corresponding email: mszliddy@uwaterloo.ca}

\begin{document}

\maketitle


\begin{abstract}
Nitrogen Vacancy Center Ensembles are excellent candidates for quantum sensors due to their vector magnetometry capabilities, deployability at room temperature and simple optical initialization and readout. This work describes the engineering and characterization methods required to control all four Principle Axis Systems (P.A.S.) of NV ensembles in a single crystal diamond without an applied static magnetic field. Circularly polarized microwaves enable arbitrary simultaneous control with spin-locking experiments and collective control using Optimal Control Theory (OCT) in a (100) diamond. These techniques may be further improved and integrated to realize high sensitivity NV-based quantum sensing devices using all four P.A.S. systems.
\end{abstract}

\newpage
\section{Introduction}\label{sec1}

Nitrogen Vacancy (NV) Centers have great potential in the area of quantum sensing. The NV's sensitivity to magnetic fields combined with their ability to be used at room temperature make them excellent test beds for exploring the engineering requirements of quantum sensing \cite{10}. Sensing applications with NV centers include imaging small magnetic fields \cite{91}, imaging nearby bacteria and molecules \cite{34,51,57,75}, sensing DC and AC magnetic fields \cite{10,1}, and sensing crystal strain in the diamond lattice \cite{70,71}. These applications may be enhanced by using ensembles of NV centers that increase the signal to noise ratio by having more active centers in the same focal volume and allow for richer sensor information to be extracted through vector measurements. \\

Control of all four NV orientations present in an ensemble, both sequentially and simultaneously, has been achieved with the use of multiple central microwave frequencies for vector magnetometry and for detecting temperature and magnetic fields simultaneously \cite{97,120,123}. This paper presents an Optimal Control Theory (OCT) controls-based solution for distinguishable manipulation of all four orientations while maintaining a compact hardware design. OCT has been previously used in NV ensembles to develop pulses robust to the nitrogen hyperfine coupling and inhomogeneities from the microwave field from control striplines; along with several other examples \cite{86, 89}. \\

Our OCT solutions are implemented using a single central control frequency and circularly polarized microwave fields, which enables control in zero applied magnetic field over a selected focal volume of NVs taken from the uniformly distributed centers in the diamond. The complex structure of the microwave control field is described by a simple Hamiltonian used to optimize OCT pulses for two key target examples: (100) and (110) diamond. Measurements were explicitly run on a (100) diamond sample to characterize phenomenological Hamiltonian parameters in order to demonstrate orientation-selective spin-locking and a set of OCT pulses that implement identity and transition-selective $\pi$ operations, respectively, over all NV orientations in the ensemble.

\section{Modelling the NV Ensemble}\label{sec2}

\subsection{NV Ensemble Structure and Circularly Polarized Microwave Control}

An NV center is created in a diamond lattice by replacing a carbon atom with a nitrogen and removing an adjacent carbon, \cite{92}. The six electrons found within the NV$^{-}$ center combine to form an effective spin-1 particle \cite{92,30,67} quantized by a zero field splitting (ZFS) $(\Delta\approx2.87$~GHz) \cite{56,81}. Generally, an NV Hamiltonian will also include a transverse ZFS due to crystal strain, $E(S_x^2 - S_y^2)$, an electron Zeeman interaction, $\gamma_e\vec{B}\cdot \vec{S}$, nuclear Zeeman interactions for both the nitrogen and nearby carbon-13 nuclei, $\gamma_{N/C} \vec{B}\cdot \vec{I_{N/C}}$, nitrogen and carbon hyperfine interactions, $A_{N/C} \vec{S}\cdot\vec{I_{N/C}}$, and a nitrogen nuclear quadrupole interaction, $QI_{Nz}^2$. However, the experiments in this work are performed in zero applied static magnetic field and in a low strain crystal so, for simplicity, the NV Hamiltonian is reduced to only the axial ZFS term: 
\begin{equation} 
\begin{split}
\CMcal{H}_{int} &= \Delta S_z^2
\label{eq:Ham_int}
\end{split}
\end{equation}

\begin{figure}[H] 
    \centering
    \includegraphics[width=0.8\textwidth]{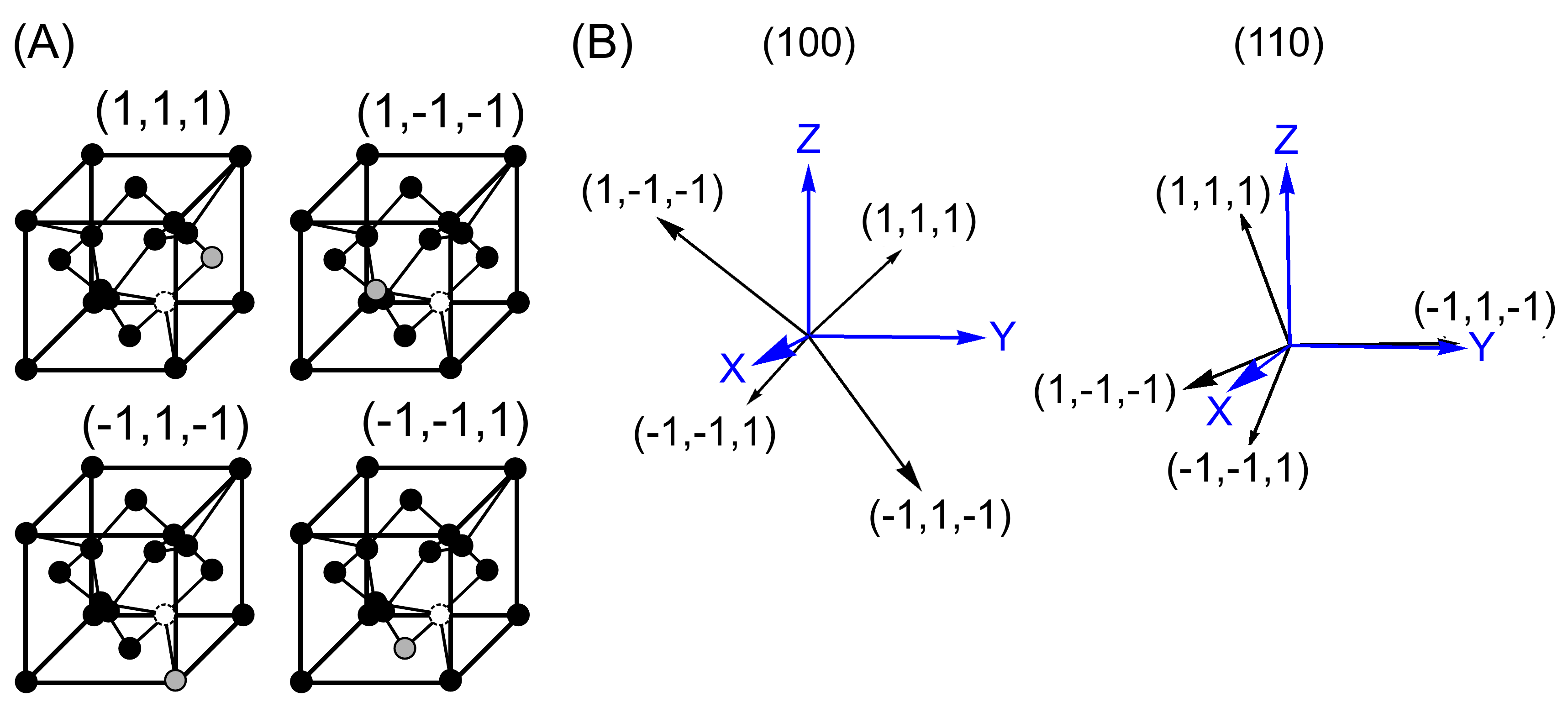}
  \caption[Caption Summary]{
  \textbf{(a)} The four possible principal axis systems (P.A.S.s) of an NV center in the diamond lattice, defined by the vector joining the Nitrogen \textbf{(grey)} and Vacancy \textbf{(white-dashed)}. The magnetic distinguishability of the four orientations make control of ensembles of NVs non-trivial. \\
  \textbf{(b)} The four P.A.S.s of NVs in the lab frame for (100) and (110) diamond. \\ Images generated in Inkscape.}
   \label{fig:1}
\end{figure}

Within the single crystal diamond structure, there are four unique possible orientations of the bond between the nitrogen and vacancy, leading to four principal axis systems (P.A.S.) for the NV Hamiltonian (Fig \ref{fig:1}). Fixing the vacancy to the relative center of the $sp_3$ hybridized structure, the orientations correspond to the Nitrogen replacing any of the four connecting carbons \cite{92,26,32,82}. Control of all four orientations simultaneously is challenging, as a single control field will not have identical action on all orientations \cite{92,56}.

The spatial orientation of the microwave field relative to the four P.A.S.s may be chosen to uniquely define either two or four sub-ensembles. For example, if the microwave field is chosen to be along the ``xz" lab plane, then it projects onto the (100) diamond to create two effective sub-ensembles: Pair A, given by the degenerate (-1,-1,-1) $\&$ (1,-1,-1) P.A.S.s; and Pair B, given by the degenerate (1,1,1) $\& $(-1,1,-1) P.A.S.s. The (100) diamond was chosen for our demonstration experiments because of this symmetry, allowing for convenient collective control and equivalent fluorescence from all orientations. This same field will project onto four unique control Hamiltonians for the (110) diamond, enabling further control application targets. \\

The conventional method for obtaining universal control of the spin-1 NV is to add a static external magnetic field that breaks the degeneracy of the $\ket{\pm1}$ ground spin states, resulting in unique splittings of the $\ket{0} \leftrightarrow \ket{+1}$ and  $\ket{0} \leftrightarrow \ket{-1}$ transitions for each P.A.S. \cite{6,60,77,121,122,127,128}. Alternatively, circularly polarized microwave fields realized with two or more channels with independent amplitude and phase control may be used to obtain transition-selective control to avoid the hardware complexity of adding an external field to the experimental setup, \cite{6,60,77,121,122,127,128}.  

\subsection{The Spatial Components of the Microwave Control Field}

\begin{figure}[H] 
\centering
\includegraphics[width=0.8\textwidth]{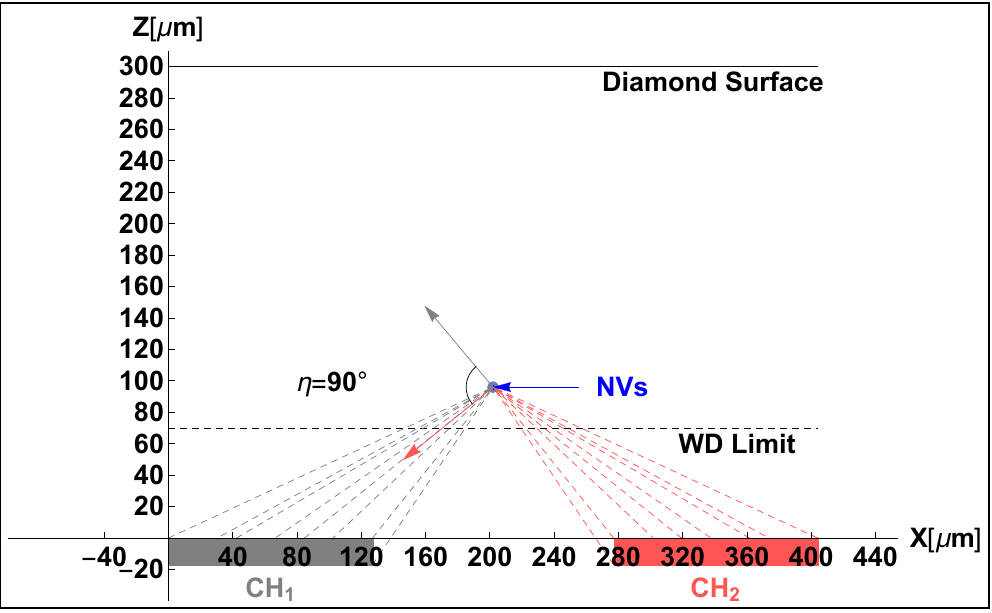}
\caption[Caption Summary]{Field distribution from each microstrip channel, CH$_1$ \textbf{(gray)}, and CH$_2$ \textbf{(red)}, at the location of the NV focal volume \textbf{(blue arrow)}, with $\eta$ indicating the orthogonality of the two fields emitted from each microstrip. The field seen by an NV at a given location is the superposition of the fields from each microstrip. The y-direction is omitted, as the microstrips are effectively infinite in that direction. The cross section of the $300~\mu$m thick diamond is shown, which combined with the $160~\mu$m working distance of the objective, places the focal volume $\approx 70\mu m$ away from the microstrips (WD limit), reducing the reflected light from the microstrips. Image generated in Mathematica.}
\label{fig:2}
\end{figure}

We chose to use two parallel microstrip resonators to create independently controllable microwave fields (figure \ref{fig:4SI}). Each microstrip was 7.5~mm long, 127~$\mu$m wide, and 17.5~$\mu$m tall, with a spacing of 150~$\mu$m between them to avoid optical interference while still delivering sufficient microwave power within the focal volume. Figure \ref{fig:2} shows the resulting microwave fields of the resonator arrangement within the cross section of a $300~\mu$m thick diamond mounted atop the two microstrips. The limit of the working distance (WD limit) of the (100x) optical objective from the top of the diamond is indicated on the figure. The $160~\mu$m working distance results in the focal area being $70~\mu$m away from the microstrips, minimizing the reflected signal of incoming green light off the microstrips. \\

 For optimal control efficiency, the orthogonality between the two fields should be maximized, $\eta = 90^{\circ}$, shown by solving for $\eta$ in the control Hamiltonian in equation \ref{eq:Ham_eff}. However, even with sub-optimal field orthogonality, OCT pulses may still be found that achieve good control outcomes, as long as there is some non-commutativity of the control fields, \cite{60,77,128}. 

The following microwave field components may be substituted into the control Hamiltonian in equation \ref{eq:Ham_eff}:
\begin{equation}
\begin{split}
w_{x1} & = \frac{\mu_o I_{1o}}{2 \pi L} \left(\arctan{\left(\frac{x-L}{z}\right)}-\arctan{\left(\frac{x}{z}\right)}+\arctan{\left(\frac{x-L}{z+h}\right)}-\arctan{\left(\frac{x}{z+h}\right)}\right) \\
w_{y1} & = 0 \\
w_{z1} & = \frac{\mu_o I_{1o}}{4 \pi L} \left(\ln{\left(\frac{z^2+x^2}{z^2+(x-L)^2}\right)}+\ln{\left(\frac{(z+h)^2+x^2}{(z+h)^2+(x-L)^2}\right)} \right)\\
w_{x2} & = \frac{\mu_o I_{2o}}{2 \pi L} \bigg(\arctan{\left(\frac{L+w-x}{z}\right)}-\arctan{\left(\frac{2L+w-x}{z}\right)} + \dots \\
& \arctan{\left(\frac{L+w-x}{z+h}\right)}-\arctan{\left(\frac{2L+w-x}{z+h}\right)}\bigg) \\
w_{y2} & = 0 \\
w_{z2} & = \frac{\mu_o I_{2o}}{4 \pi L} \left(\ln{\left(\frac{z^2+(L+w-x)^2}{z^2+(2L+w-x)^2}\right)}+\ln{\left(\frac{(z+h)^2+(L+w-x)^2}{(z+h)^2+(2L+w-x)^2}\right)} \right)
\end{split}
\end{equation}
Where $\mu_o$ is the magnetic permeability for free space, $I_{1/2o}$ is the current running through each microstrip, $L$ is the length of the microstrips, $w$ is the distance between each microstrip and $x$ and $z$ are the location of the center of the focal volume relative to the origin. \\

\newpage
\subsection{P.A.S.-Dependent control Hamiltonian}

There are three important frames to consider when describing the control Hamiltonian: the lab frame, the diamond crystal frame, and the NV P.A.S.. Expressing all components in a common frame, and then rotating into the frame of the internal Hamiltonian (eq.\ref{eq:Ham_int}) yields the control Hamiltonian in equation \ref{eq:Ham_eff}, the full derivation of which may be found in the supplementary information: \\
\begin{equation}
\begin{split}
\CMcal{H}_{ctrl} & = A S_x + B S_y' + C S_y + D S_x' \\
& A = \frac{1}{2}(R_{xx} (I_1w_{x1} +I_2w_{x2})+ R_{yx} (I_1w_{y1} +I_2w_{y2}) + R_{zx} (I_1w_{z1}+I_2 w_{z2})) \\
& B = - \frac{1}{2}(R_{xx} (Q_1w_{x1} +Q_2w_{x2})+ R_{yx} (Q_1w_{y1} +Q_2w_{y2}) + R_{zx} (Q_1w_{z1}+Q_2 w_{z2})) \\
& C =  \frac{1}{2}(R_{xy} (I_1w_{x1} +I_2w_{x2})+ R_{yy} (I_1w_{y1} +I_2w_{y2}) + R_{zy} (I_1w_{z1}+I_2 w_{z2})) \\
& D = - \frac{1}{2}(R_{xy} (Q_1w_{x1} +Q_2w_{x2})+ R_{yy} (Q_1w_{y1} +Q_2w_{y2}) + R_{zy} (Q_1w_{z1}+Q_2 w_{z2})) 
\end{split}
\label{eq:Ham_eff}
\end{equation}
The Hamiltonian is expressed in a letter format to show the division between the four standard spin-1 operators, $S_{x/y}$, and ``twisted" spin-1 operators, $S_{x/y}^{'}=i[S_{y/x},S_z^2]$, resulting from moving into the interaction frame of the internal Hamiltonian \cite{43,thesis}. It is the presence of all four of these operators that allows for access to transition-selective control in the absence of an applied static magnetic field. This Hamiltonian displays a dependence on the geometric relationship of the NV P.A.S.s ($R_{xx},R_{xy},R_{zx}$ etc.(SI Fig: \ref{fig:7SI})) unique to each NV, the spatial components of the microwave control field ($w_{x1},w_{x2}$ etc.), and the four control channels of an AWG ($I_1,I_2,Q_1,Q_2$) used to create the control signals. \\

\section{OCT Design of Transition-Selective Pulses in Zero Field}

Optimal control theory (OCT) is used extensively in quantum control and has recently been proven useful for finding robust control solutions in quantum sensing \cite{86,89,84}. OCT algorithms generally proceed by optimizing a set of parameterized controls to achieve a desired quantum operation subject to constraints when calculated over a set of Hamiltonians and noise/environment processes. In the most common implementations, the quantum operation is defined to specify either desired state-to-state transitions \cite{REF1, REF2} or a full unitary or completely positive trace-preserving (CPTP) map \cite{REF6}. Noise/environment processes and constraints can generally include any process that can be modelled compactly, such as field inhomogeneities \cite{REF3, REF10}, limited Rabi frequency \cite{REF9}, and control system distortions \cite{REF4, REF5}. \\

In our control situation, the projection of the control fields onto the unique set of NV P.A.S. orientations leads to an incoherent distribution of Hamiltonians that may be treated by optimizing over a direct sum representation of the dynamics \cite{REF3, REF7, REF8}. The validity of the direct sum representation is helped by the average dipolar coupling between NV centers in the chosen experimental sample being on the order of kilohertz, allowing the control Hamiltonians to be considered independent over each subensemble. Coupling between NV centers may be added for future iterations \cite{115}. \\ 
 
In our application, OCT pulses are found using the gradient ascent pulse engineering (GRAPE) algorithm implemented in the Quantum Utils package \cite{89,43,thesis,52,114}. To create a pulse, an initial guess of the pulse shape is made, then updated successively using gradient techniques until the desired map or state-to-state transfer is achieved up to a target performance. The target unitary or state-to-state transfer is given a target fidelity or state overlap (0.99), internal Hamiltonian ($\Delta S_z^2$), set of control Hamiltonians (eq.\ref{eq:Ham_eff}) and parameterized controls ($I_1,I_2,Q_1,Q_2$) as inputs. The total length of the pulse is given by the number of time steps multiplied by the length of each time step. The chosen physical control fields result in two or four control Hamiltonians for the (100) and (110) diamonds, respectively. \\

\subsection{Defining Transition-Selective Maps}

To aid finding and visualizing intuitive solutions, pseudo spin-$\frac{1}{2}$ operators may be used in place of standard spin-1 operators \cite{Abragam}. These are not true spin-$\frac{1}{2}$ operators, as the $\ket{\pm 1}$ states share a space with the same $\ket{0}$ state, but conveniently represent the pulse action over states of interest. The pseudo spin-$\frac{1}{2}$ operators are labelled as $S_{x/y/z}^{\pm}$ for the $\ket{0},\ket{+1}$ and $\ket{0},\ket{-1}$ state pairs, respectively. Expressions for pseudo spin-$\frac{1}{2}$ operators in terms of spin-1 operators are shown below, with the matrix forms found in equation \ref{eq:pseudo_spin_1_2_operators} of the supplementary information.
\begin{equation}
S_x^{\pm} = \frac{1}{\sqrt{2}}(S_x \pm i[S_y,S_z^2])
\end{equation}
\begin{equation}
S_y^{\pm} = \frac{1}{\sqrt{2}}(S_y \mp i[S_x,S_z^2])
\end{equation}
\begin{equation}
S_z^{\pm} = \frac{1}{2} S_z \pm \mathds{1}_3 \mp \frac{3}{2} S_z^2
\end{equation} 
The pseudo spin-$\frac{1}{2}$ operators obey the standard commutation relations of the Pauli operators (i.e. $[S_x^{\pm},S_y^{\pm}]=2iS_z^{\pm}$), allowing for maps to be constructed in the same fashion as for spin-$\frac{1}{2}$ particles. As such, the OCT maps may be defined with the general unitary operator in equation \ref{eq:U}, where ($\alpha$) is the rotation angle, $(\hat{n})$ is the unit vector defining the rotation axis, and $S^{\pm}$ defines the total list of operators, $S^{\pm}=\{Sx^{\pm},Sy^{\pm},Sz^{\pm}\}$. The (+) or (-) operators are chosen depending on which state transitions are desired, the matrix form of the positive operator shown as an example.  
\begin{equation}
\begin{split}
U^{\pm} & = \mathds{1}_3-\mathds{1^{\pm}}+\cos{\left(\frac{\alpha}{2}\right)}\mathds{1}^{\pm}-i\sin{\left(\frac{\alpha}{2}\right)} \left(\hat{n}\cdot S^{\pm}\right) \\
U^{+} & = \left(
\begin{array}{ccc}
   1 & 0 & 0\\
   0 & -\cos{\left(\frac{\alpha}{2}\right)} + in_z\sin{\left(\frac{\alpha}{2}\right)} & (in_x+ n_y)\sin{\frac{\alpha}{2}} \\
   0 & (in_x - n_y)\sin{\frac{\alpha}{2}} & -\cos{\left(\frac{\alpha}{2}\right)} - in_z\sin{\left(\frac{\alpha}{2}\right)}
\end{array}
\right)
\end{split}
\label{eq:U}
\end{equation}

\subsection{Transition-Selective OCT Pulse}

\begin{figure}[H] 
\centering
\subfigure{\includegraphics[width=1\textwidth]{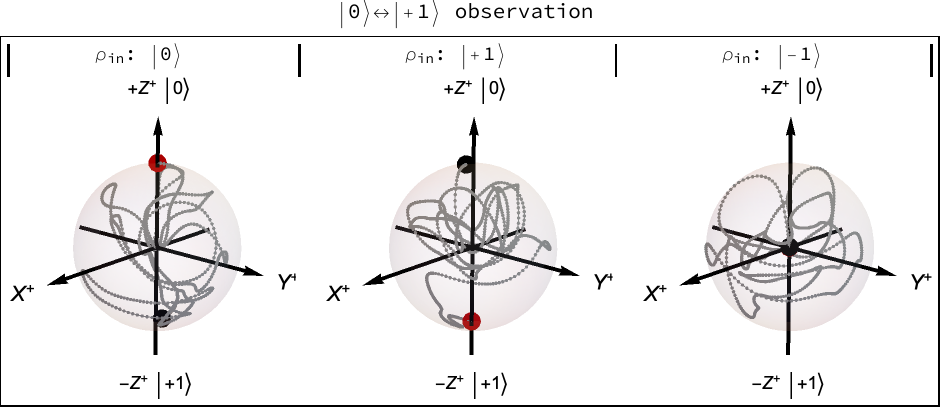}}
\subfigure{\includegraphics[width=1\textwidth]{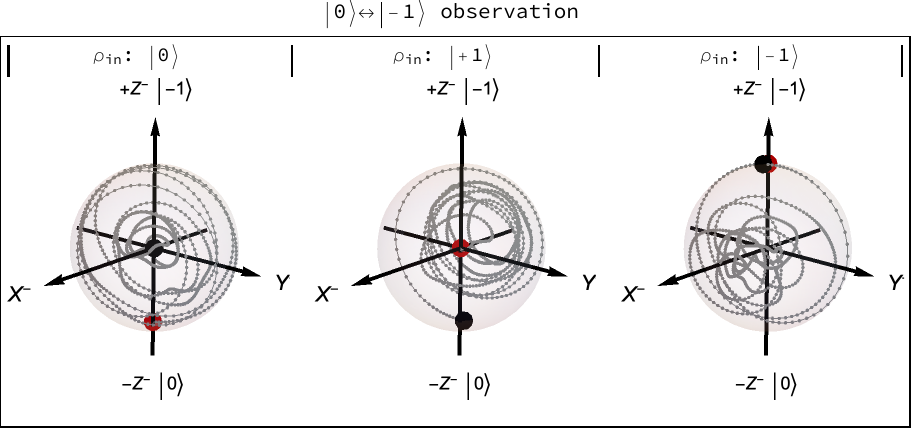}}
\caption[Caption Summary]{Bloch sphere representations of the action of a transition-selective OCT pulse, optimized to implement a selective $\pi_+$ operation. The ``z"-axis for the positive and negative spheres correspond to the $\ket{0}$ and $\ket{+(-1)}$ states. The positive ``x"-axis corresponds to the $\frac{\ket{0}+\ket{\pm1}}{\sqrt{2}}$ state for the positive and negative Bloch spheres, while the positive y-axis corresponds to the $\frac{\ket{0}+i\ket{\pm1}}{\sqrt{2}}$ state. The starting state is indicated as a red dot, the final state as a black dot, and the state trajectory as a grey line. In this figure, only the trajectory of the pulse on sub-ensemble (A) is shown. The OCT pulse may be seen to successfully implement a $\pi_+$ transition between the $\ket{0}$ and $\ket{+1}$ states while simultaneously performing an identity operation if the starting state is $\ket{-1}$. Images generated in Mathematica.}
\label{fig:6}
\end{figure}

As an example of optimizing a transition-selective OCT pulse, consider a selective $\pi_{+/y}$ pulse on the (100) diamond that enacts a $\pi_y$ rotation in the positive pseudo-subspace and an identity operation in the negative: ($\alpha=\pi, n_x=0, n_y=1, n_z=0$). The action of this pulse may be understood through two sets of Bloch spheres, defined over $S^{\pm}$, respectively through $\left<(X/Y/Z)^{+/-}\right> = Tr[S_{(x/y/z)}^{+/-}.\rho_t]$ \cite{thesis}. Figure \ref{fig:6} shows the action of a single OCT pulse on the positive \textbf{(top)} and negative \textbf{(bottom)} Bloch spheres, respectively. The starting state is indicated with a \textbf{red} sphere and final state with a \textbf{black} sphere, while the trajectory of the pulse is shown in grey. \\

Recall that the (100) diamond has only two non-degenerate elements in the incoherent distribution of control Hamiltonians under the chosen microwave field orientation. The trajectory for the (-1,-1,1) $\&$ (1,-1,-1) NV sub-ensemble (A) has been shown. The trajectory of the (1,1,1) $\&$ (-1,1,-1) NV sub-ensemble (B) is shown in figure \ref{fig:6SI}. The mapping of the selective $\pi_{+}$ pulse is clearly seen with the starting states of $\ket{0}$ and $\ket{+1}$ in the positive Bloch spheres, rotating to $\ket{+1}$ and $\ket{0}$, respectively. The shared $\ket{0}$ state may also be observed with the negative Bloch sphere. The correct operation of the intended identity pulse, beginning in the $\ket{-1}$ state is most clearly observed in the negative Bloch sphere. This is further reflected in the positive sphere as the starting and ending state is observed to be at the origin of the positive sphere, indicating no population. 

\subsection{Orientation-Selective OCT Pulse}

\begin{figure}[H] 
\centering
\subfigure{\includegraphics[width=0.9\textwidth]{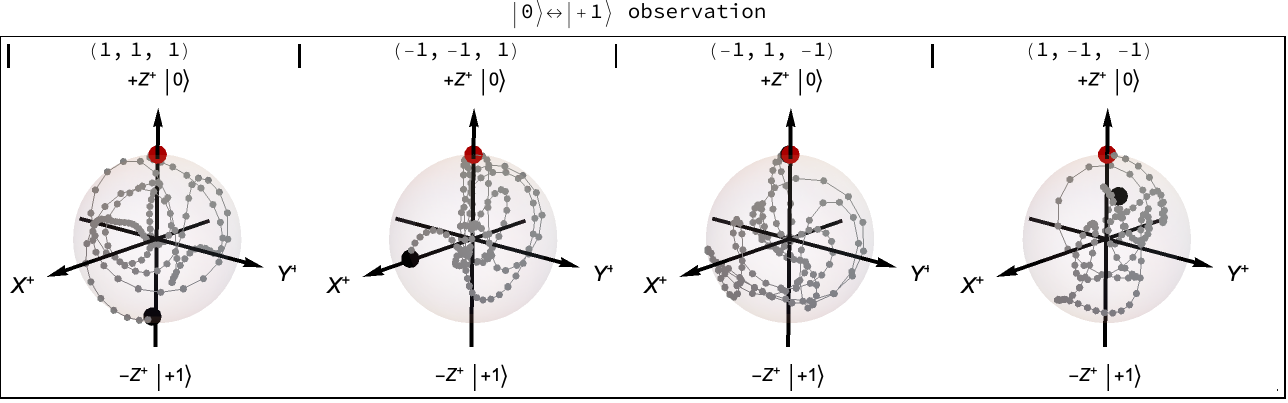}}
\subfigure{\includegraphics[width=0.9\textwidth]{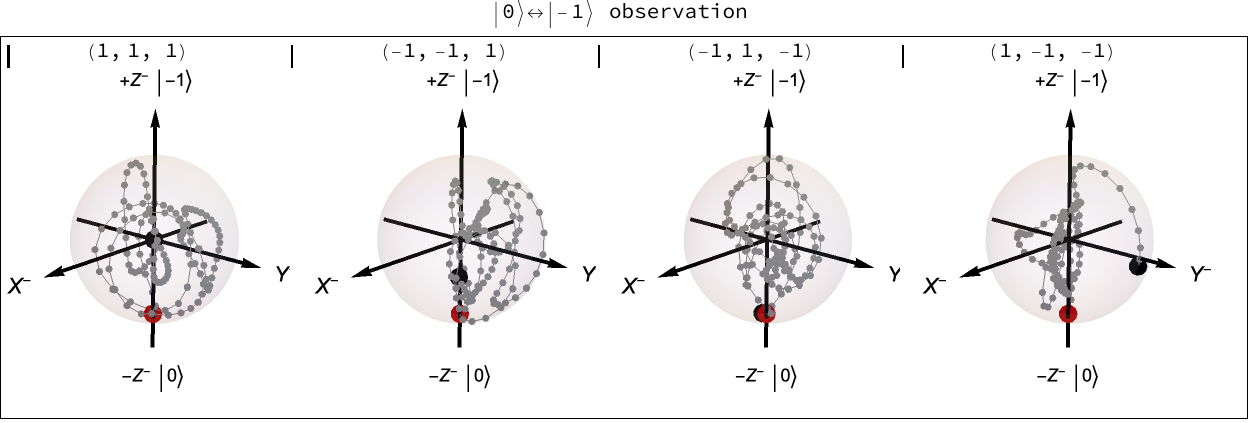}}
\caption[Caption Summary]{Simulation of an OCT pulse that implements a unique operation on each NV orientation within a (110) diamond. The simultaneous success of each operation showcases the ability for OCT pulses to perform both transition-selective and orientation-selective operations within a single pulse. Images generated in Mathematica.}
\label{fig:110}
\end{figure}

While the (100) diamond is an excellent candidate for demonstrating collective control with two sub-ensembles of NV pairs, the (110) diamond configuration may be used to investigate the full potential of separately controlling all four unique NV orientations within the single crystal. In this case there are four non-degenerate elements in the incoherent distribution of control Hamiltonians under the chosen microwave field orientation. Figure \ref{fig:110} shows the Bloch sphere representations of the action of a single orientation-selective OCT pulse (Table \ref{table:1}) that simultaneously implements a unique operator for each NV orientation.
\begin{table}[h]
\begin{center}
\begin{minipage}{174pt}
\caption[Operators for each NV Orientation]{Operators for each NV Orientation in figure \ref{fig:110}}\label{table:1}
\begin{tabular}{@{}cc@{}}
\hline
Operator & NV Orientation  \\ 
\hline
$\pi)_{x/+}$ & (1,1,1) \\ [1ex]
 \hline 
 & \\ [-0.5ex]
 ~$\frac{\pi}{2}\big)_{\overline{y}/+}$ & (-1,-1,1) \\ [2ex]
\hline
 & \\ [-0.5ex]
 ~$\mathds{1}_{\pm}$ & (-1,1,-1) \\ [1ex]
 \hline
  & \\ [-0.5ex]
 $\frac{\pi}{2}\big)_{\overline{x}/-}$ & (1,-1,-1) \\ [1ex]
\hline
\end{tabular}
\end{minipage}
\end{center}
\end{table}

\section{Characterization of the NV Ensemble}

\subsection{The Sample and Experimental System Design} 

For all the experiments performed, a $500~\mu$m thick DNV-B1 (100) diamond from Element Six was used, which contained an estimated 16~000 NV centers within a focal volume of beam diameter 0.59~$\mu$m and depth of field 2.51~$\mu$m \cite{99,100}. This sample was chosen, as opposed to the ideal $300~\mu$m thick sample, to keep the focal volume well out of range of the microstrips, removing any reflected light from the microstrips. The Gaussian optics at the site of the focal volume are shown in figure \ref{fig:1SI} in the supplementary information. Extending beyond the sample, a description of the full optical layout may also be found in the supplementary information. As the experiments were performed at room temperature, the NVs were excited with off-resonant green light, and the red light emission collected with an avalanche photo-diode (APD). 

\subsection{First Calibration Experiment - Equal Fluorescence from All Orientations}

To remove any bias towards any one orientation, the fluorescence from each orientation first had to be equalized. The orientation-dependent fluorescence from the centers may be controlled by changing the incoming optical polarization \cite{thesis,5,73,110}. Equation \ref{eq:int} describes how the output intensity of the NV is related to how much the P.A.S. ``z"-axis deviates from the propagation axis of the incoming light, ($\theta$), and how much the NVs' ``x"-axis deviates from the polarization of the incoming light, ($\phi$). The total emission is the sum of the emission from the $E_y$ and $E_x$ excited states, where maximizing $E_y$ minimizes $E_x$, and vice-versa. $E_y$ is dependent only on the $\phi$ angle, while $E_x$ is dependent on both $\phi$ and $\theta$ \cite{thesis,5,73,110}:
\begin{equation}
\begin{split}
IE_y(\phi) & = \sin(\phi)^2 \\
IE_x(\theta, \phi) & = \cos(\theta)^2\cos(\phi)^2  \\
ITot(\theta,\phi) & = \sin(\phi)^2 + \cos(\theta)^2\cos(\phi)^2 
\end{split}
\label{eq:int}
\end{equation}

\begin{figure}[H] 
\centering
\includegraphics[width=0.8\textwidth]{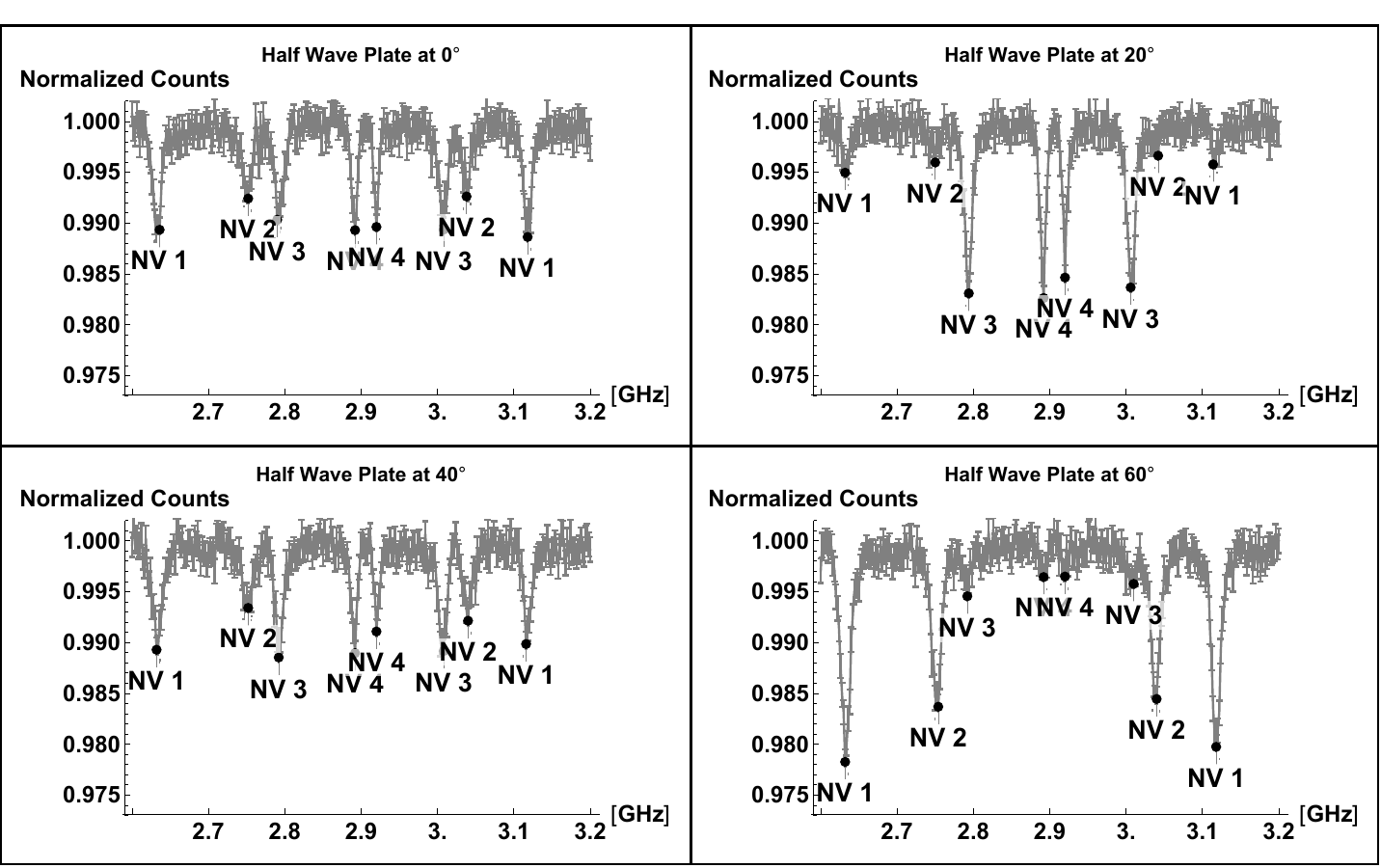}
\caption[Caption Summary]{ODMR-CW spectra as a function of incoming optical polarization, dictated by the rotation of a HWP in Degrees. A static magnetic field was added to resolve eight peaks from the four orientations. Changes in incoming light polarization change the alignment of the light electric field with the electric dipole of the NV, causing fluorescence to be favoured in some orientations. For (100) diamond,a single polarization value could be found with equal fluorescence from each NV orientation, removing bias in the control measurement outcomes.}
\label{fig:3}
\end{figure}

(100) diamond is an experimentally convenient crystal as all NV orientations deviate from the light propagation axis at the same angle ($\theta$), so fluorescence may be optimized by adjusting only the $\phi$ parameter. Figure \ref{fig:3} shows the optically detected magnetic resonance-continuous wave (ODMR-CW) spectra of the diamond with a static magnetic field added for convenience to resolve the four orientations into eight peaks. ODMR-CW experiments were performed to monitor relative fluorescence incrementally as a Half Wave Plate (HWP) was rotated, beginning at a relative $0^{\circ}$. At $20^{\circ}$ and $60^{\circ}$, the fluorescence favouring either of the two orientations may be seen. At $0^{\circ}$ and $40^{\circ}$, a more even emission from each orientation is indicated. The optimal value providing even fluorescence across all orientations was found to be $42^{\circ}$.

\subsection{Determining a Phenomenological Control Hamiltonian}

Equation \ref{eq:Ham_eff} provides a formal description of the influence of the P.A.S., microwave spatial components, and AWG controls on the form of the control Hamiltonian. However, accurately determining the parameters necessary to fully specify the formal Hamiltonian proved difficult. Instead, the OCT experiments performed in this work enlist a phenomenological control Hamiltonian (equation \ref{eq:Ham_phen}), that includes experimentally measured parameters. The two microstrips provide the input control amplitudes ($\Omega_{1/2}$) and global control phase ($\Delta\theta$), recalling these are simply the polar coordinates of the Cartesian controls shown in the \emph{a priori} Hamiltonian. The Hamiltonian also contains the measured Rabi drive strength $(\Omega_{NV_{A/B}})$ for each of the NV sub-ensembles pair (A) and (B) and the experimental phase value $(\eta)$ \cite{thesis}. 
\begin{equation}
\begin{split}
\CMcal{H}_{A/B} & =  \Omega_{NV_{A/B}} (\frac{\Omega_1}{2}S_x + \frac{\Omega_2}{2} \cos(\Delta \theta)\cos(\eta)S_x + \frac{\Omega_2}{2} \cos(\Delta \theta)\sin(\eta)S_y \\
& + \frac{\Omega_2}{2} \sin(\Delta \theta)\sin(\eta)S_x^{'} + \frac{\Omega_2}{2} \sin(\Delta \theta)\cos(\eta)S_y^{'})
\label{eq:Ham_phen}
\end{split}
\end{equation}
    
\begin{figure}[H] 
\centering
\includegraphics[width=0.8\textwidth]{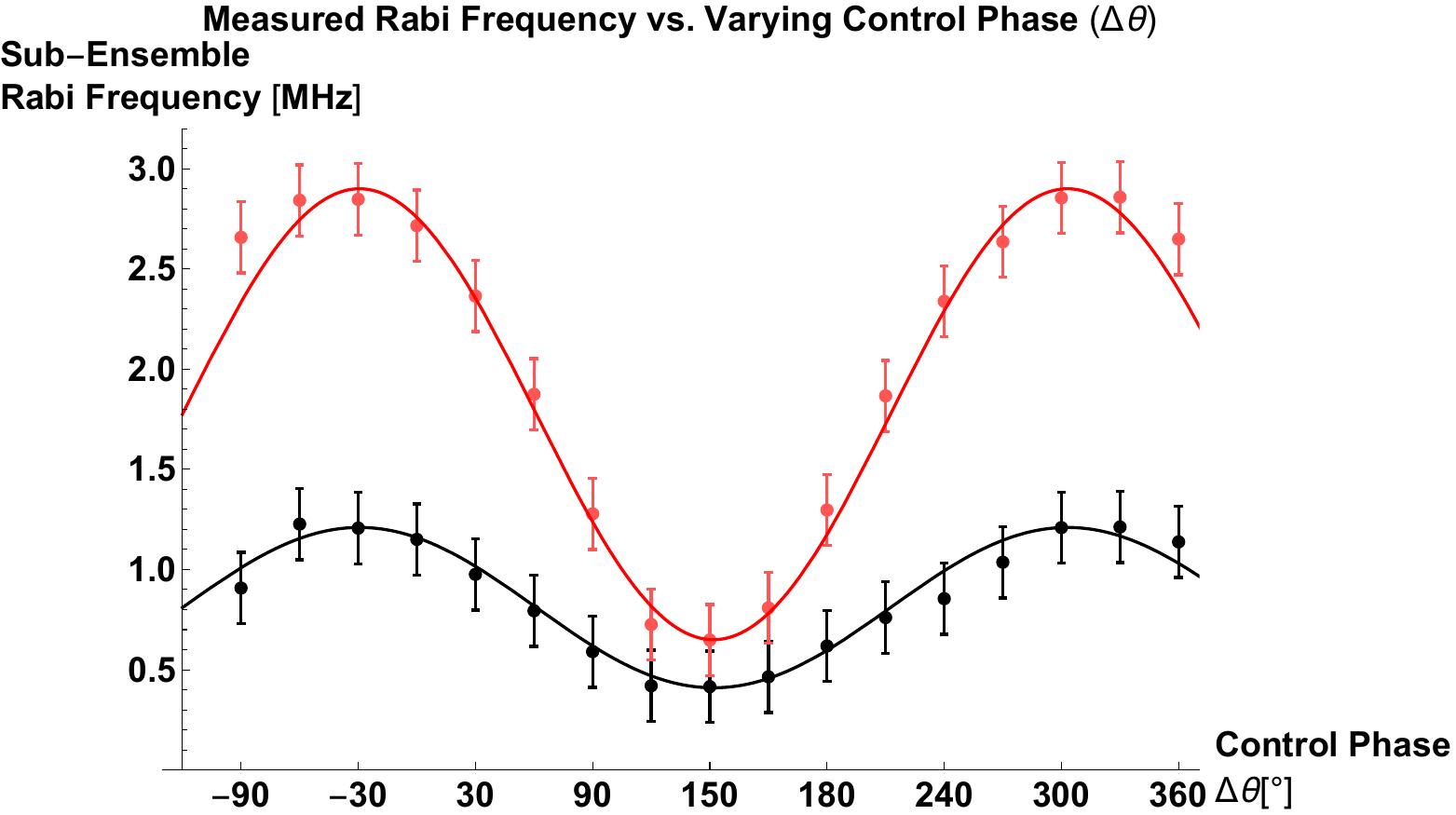}
\caption[Caption Summary]{The Dual Channel Rabi Experiment shows the sub-ensemble Rabi frequency dependence on the relative control phase. The high frequency sub-ensemble (A) (-1,-1,-1)$\&$(1,-1,-1) is indicated with \textbf{(red)} and low frequency sub-ensemble (B) (1,1,1)$\&$(-1,1,-1), with \textbf{(black)}. For each data point, the relative phase was changed between the two channels $\Delta\theta$ and resulting Rabi frequencies $\Omega_{NV_{A/B}}$ gathered. The results yield a very strong phase dependence on the measured Rabi frequency for each of the sub-ensembles as a function of $\cos(\Delta\theta)^2$. For our OCT experiments, the control phase values were fixed, and amplitude-only control was used for pulse design.} 
\label{fig:5}
\end{figure}

A Dual Channel Rabi experiment was used to measure the strength of the Rabi drive for each sub-ensemble, $\Omega_{NV_{A/B}}$. Figure \ref{fig:5} shows how $\Omega_{NV_{A/B}}$ (``y-axis") changes by varying the control phase $\Delta\theta$. In this case, the phase of the second channel varied $\theta_2=\Delta\theta$ (``x-axis") while maintaining a fixed phase of the first channel $\theta_1=0^{\circ}$ and fixed control amplitude $\Omega_{1/2}=1$ \cite{60, thesis}. The error bars indicate the full width at half maximum of the gathered spectra.

The results show a strong dependence of the measured Rabi frequencies $\Omega_{NV_{A/B}}$ on the relative phase $\Delta\theta$ as a $\cos(\Delta\theta)^2$ function. The phase response is more dramatic for the high-frequency (-1,-1,-1)$\&$(1,-1,-1) pair (A) \textbf{(red)} than for the low-frequency (1,1,1)$\&$(-1,1,-1) pair (B) \textbf{(black)}. To accommodate the relationship between the input control phase and output $\Omega_{NV_{A/B}}$, a map associating this behaviour would have to be included in optimization for each input control phase ($\Delta\theta$). Our chosen alternative approach was to fix the input control phase, resulting in a fixed $\Omega_{NV_{A/B}}$, and only optimize the two input amplitudes ($\Omega_{1/2}$) using OCT. 

\section{Demonstrating Control of NV Ensembles}

\subsection{Demonstrating Orientation Selectivity With a Spin-locking Experiment} 

\begin{figure}[H]
\centering
\includegraphics[width=0.5\textwidth]{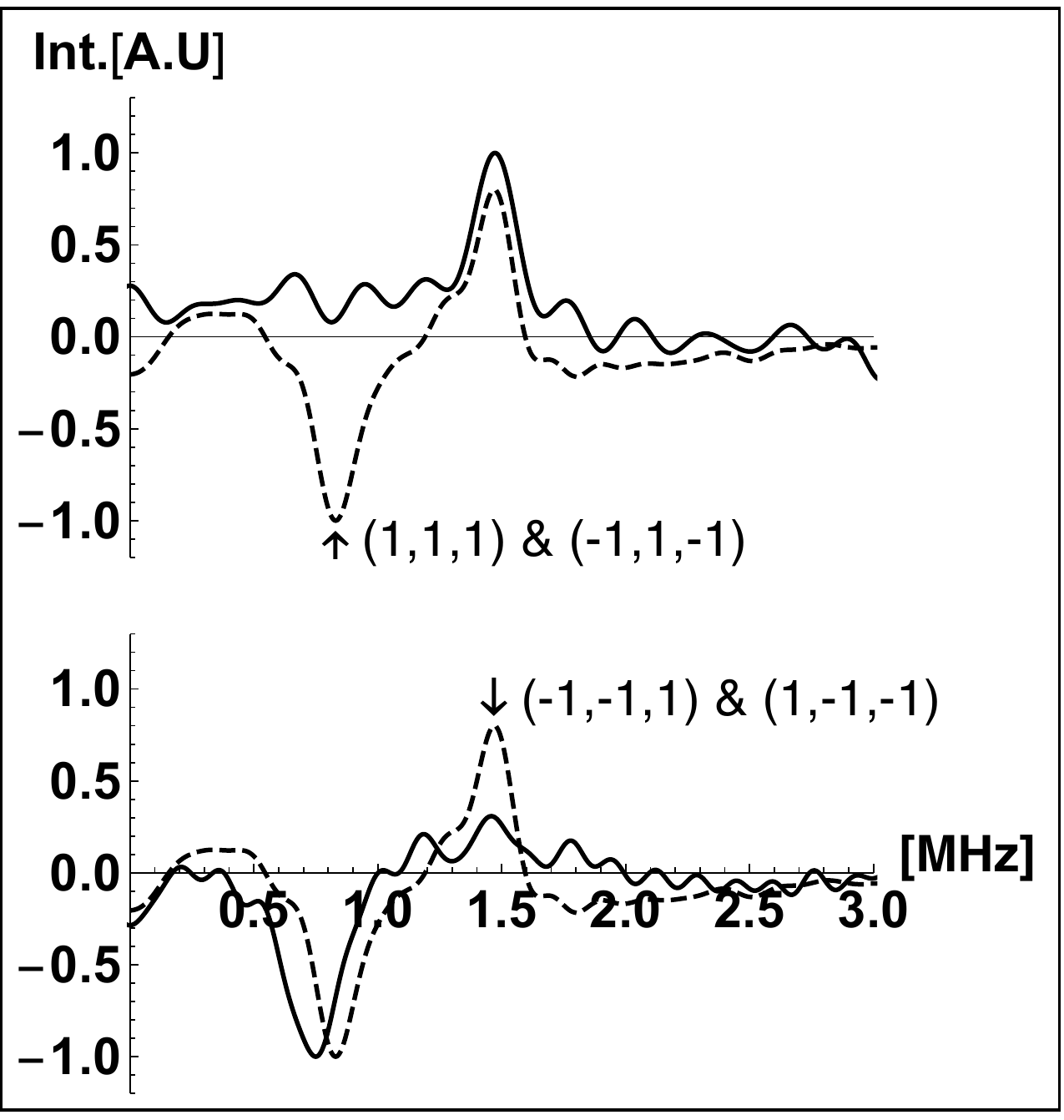}
\caption[Caption Summary]{The Rabi frequency spectra \textbf{(black, dashed)} show two frequencies from the two sub-ensembles of NV orientations in the (100) diamond. The spin-locking results are overlaid \textbf{(black, solid)}, indicating which of the two sub-ensemble is being suppressed in the experiment with an arrow. There is a clear selective suppression of either the low- or high-frequency peak for each experiment. The spin-locking experiment provides a controls-based solution comparable to the results obtained when optical polarization is changed, suppressing the fluorescence of a chosen orientation.}
\label{fig:4}
\end{figure}

Preliminary control was demonstrated with a spin-locking experiment that suppresses the evolution of each sub-ensemble selectively \cite{thesis}. Only one microwave channel is required to perform the spin-locking experiment. In this experiment, the control Hamiltonian rotates one sub-ensemble to the "x``-axis, suppressing its evolution while allowing the other sub-ensemble to evolve, the details of which are in the supplementary information (figure \ref{fig:5SI}). The results of the spin-locking experiment \textbf{(black, solid)} are plotted over the Rabi spectra \textbf{(black,dashed)} in figure \ref{fig:4}, with arrows indicating which sub-ensemble was suppressed. 

There is a clear selective suppression for each of the sub-ensembles in the spin-locking experiment. The ability to suppress the signal of each of the sub-ensembles allows for each to be studied independently, providing a controls-based solution as an alternative to changing the optical polarization to suppress fluorescence from individual orientations.

\subsection{Experimental Implementation of OCT}

\begin{figure}[H] 
\centering
   \includegraphics[width=0.75\textwidth]{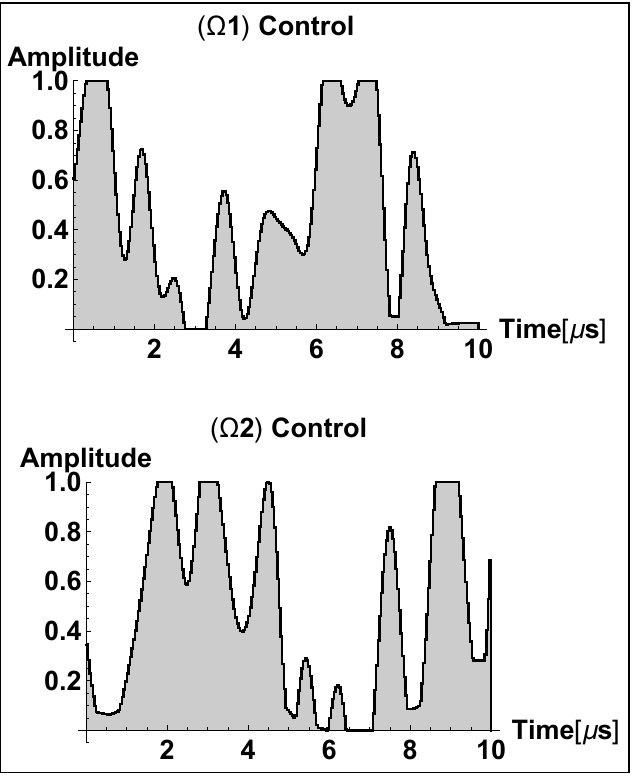}
   \caption[]{Sample pulse, showing the optimizing over the control amplitudes $\Omega_{1/2}$ to create an OCT pulse. The OCT pulses accept the fixed control phase $\Delta\theta$, and experimental values $\Omega_{NV_{A/B}}$ and $\eta$ within the control Hamiltonian and optimize using the free control amplitudes $\Omega_{1/2}$ to create the pulse. Each pulse length is given by the time step multiplied by the number of steps allotted, here being 40ns and 250 steps to yield a pulse length of $10~\mu$s. To avoid hardware distortion issues, the pulses should be as smooth as possible, a key factor in deciding the length of the time step.} 
\label{fig:SI_Amp_Controls}
\end{figure}

Expanding beyond the spin-locking experiments, a few proof of concept OCT experiments were performed. The dual channel Rabi experiment yielded $\Omega_{NVA}=2.64~$MHz and $\Omega_{NVB}=1.03~$MHz, with an experimental phase value of $\eta=115^{\circ}$, at a fixed global control phase of $\Delta\theta=270^{\circ}$. These values were used to optimize OCT pulses using amplitude-only controls ($\Omega_{1/2}$ in equation \ref{eq:Ham_phen}). The resulting optimized pulse shape is shown in figure \ref{fig:SI_Amp_Controls}. Pulses were optimized using a state-to-state target similar to an Identity operation, $\ket{0}\bra{0} \rightarrow \ket{0}\bra{0}$, and selective $\pi$-type operation $\ket{0}\bra{0} \rightarrow \ket{+1}\bra{+1}$, respectively. The state-to-state transfer could be achieved in 250 steps of 40 ns each, for a total pulse length of $10~\mu$s. In contrast, 800 steps ($32~\mu$s long pulse) were necessary to achieve a complete selective $\pi_+$ map under the same experimental conditions. \\

The results of the experimental implementation of several different optimized OCT pulses are shown in table \ref{table:2} \cite{thesis}. The experimental photon counts for each pulse implementation are normalized to a reference count of the $\ket{0}$ state. In an ideal case, the Identity-like operation would yield a population of 100$\%$ $\ket{0}$ and the $\pi$-like operation, $0\%$. The results show a reasonable distinguishability between the behaviour of the two pulse types, with consistent results optimizing over either sub-ensemble (A), (B), or both.  

\begin{table}[h]
\begin{center}
\begin{minipage}{340pt}
\caption[OCT Results]{Normalized photon count collected following implementation of six OCT pulses, and the corresponding population of the $\ket{0}$ state. These experiments represent an Identity $(\ket{0}\bra{0} \rightarrow \ket{0}\bra{0})$ and single transition $\pi$ pulse $(\ket{0}\bra{0} \rightarrow \ket{+1}\bra{+1})$ optimized for NV pair (A), (B), and both pairs (AB).}\label{table:2}
\begin{tabular}{@{}ccccccc@{}}
\hline
Experiment & $\mathds{1}_{A}$ & $\pi_{+A}$ & $\mathds{1}_{B}$ & $\pi_{+B}$ & $\mathds{1}_{AB}$ & $\pi_{+AB}$ \\
\hline
Normalized Counts & 0.94(0) & 0.91(7) & 0.94(0) & 0.91(9) & 0.93(9) & 0.92(5) \\ [1.5ex]
Population of $\ket{0}$ & 40$\pm$1.5$\%$ & 17$\pm$1.5$\%$ & 40$\pm$1.5$\%$ & 19$\pm$1.5$\%$ & 39$\pm$1.5$\%$ & 25$\pm$1.5$\%$ \\
\hline
\end{tabular}
\end{minipage}
\end{center}
\end{table}

The limited Rabi drive strength for each of the NV sub-ensembles required the use of long pulses that were sensitive to experimental unknowns, limiting the contrast between the identity and $\pi$ pulses. To improve the results, another set of pulses was optimized that accounted for a small variation in the zero field splitting ($\approx 100$~kHz) while maintaining the same $10~\mu$s pulse length. These robust pulses improved the population results of the identity experiments to $53\%$, while leaving the $\pi$ pulse performance unchanged \cite{thesis}. The hyperfine interaction with the nitrogen nuclear spin was also identified as a significant source of error. Simulations of the experimentally implemented pulses that included the nitrogen hyperfine interaction gave performance that more closely matched the experimental results. Robustness to this hyperfine interaction was not included in optimizations as the control amplitude for each of the ensembles (1.03~MHz $\&$ 2.64~MHz) was too small to effectively account for the hyperfine interaction (2.16~MHz). 

\section{Conclusion}

This paper presented a controls based solution with a generalized control Hamiltonian for understanding the dynamics of an ensemble of NVs within a (100) and (110) single crystal diamond. We demonstrated primitive control of NVs within a (100) diamond using a spin-locking experiment, then using measured values gathered through characterization experiments, collective control for all orientations of NVs with a phenomenological Hamiltonian was demonstrated with distinguishable proof of concept Identity and $\pi_+$ experiments using OCT pulses. \\
    
These proof of concept experiments on the (100) diamond were completed with an arbitrary NV focal volume, minimal distinguishability between the two control fields and one central control frequency. In addition, early optimization with a small static field showed a large improvement in the pulse success, indicating the potential for success in future robustness optimizations. \\

Having an arbitrary focal volume did showcase the ability of the OCT pulses, but having some guidance on the focal volume would improve the non-commutivity between the pulses, extending their capabilities. Using a thinner diamond would allow for the NV focal volume to be closer to the microstrips, increasing the Rabi drive strength and reducing the length of the pulse, allowing for better robustness to RF inhomogeneities and hyperfine splittings with the nitrogen and nearby carbon nuclear spins. Robust OCT pulses can be incorporated into existing chemical sensing schemes to enhance their capabilities. \\

Without altering the experimental setup, a (110) diamond may be used, where four unique orientations are desired. In this case, each orientation may be selectively controlled to either sequentially or simultaneously detect complex magnetic fields as desired by the experiment, further expanding the applications of OCT controlled systems. 


\newpage
\section*{Declarations}

\subsection{Funding}

This research was undertaken thanks in part to funding from the Canada First Research Excellence Fund (CFREF) and the funding collaboration with NSERC DND. The authors declare to have no financial interests. 

\subsection{Conflict of interest}

The authors declare that they have no conflict of interest.

\subsection{Availability of data and materials}

The datasets generated during and/or analysed during the current study are available from the corresponding author on reasonable request.

\newpage
\bibliographystyle{unsrt}
\bibliography{bibliography}

\begin{thebibliography}{10}

\bibitem{10}
John~F. Barry, Jennifer~M. Schloss, Erik Bauch, Matthew~J. Turner, Connor~A.
  Hart, Linh~M. Pham, and Ronald~L. Walsworth.
\newblock Sensitivity optimization for {NV}-diamond magnetometry.
\newblock {\em Rev. Mod. Phys.}, 92(1):015004, March 2020.
\newblock Publisher: American Physical Society.

\bibitem{91}
L~Rondin, J-P Tetienne, T~Hingant, J-F Roch, P~Maletinsky, and V~Jacques.
\newblock Magnetometry with nitrogen-vacancy defects in diamond.
\newblock {\em Rep. Prog. Phys.}, 77(5):056503, May 2014.

\bibitem{34}
David~R. Glenn, Kyungheon Lee, Hongkun Park, Ralph Weissleder, Amir Yacoby,
  Mikhail~D. Lukin, Hakho Lee, Ronald~L. Walsworth, and Colin~B. Connolly.
\newblock Single-cell magnetic imaging using a quantum diamond microscope.
\newblock {\em Nature Methods}, 12(8):736--738, August 2015.
\newblock Number: 8 Publisher: Nature Publishing Group.

\bibitem{51}
Zeeshawn Kazi, Isaac~M. Shelby, Hideyuki Watanabe, Kohei~M. Itoh,
  Vaithiyalingam Shutthanandan, Paul~A. Wiggins, and Kai-Mei~C. Fu.
\newblock Wide-{Field} {Dynamic} {Magnetic} {Microscopy} {Using}
  {Double}-{Double} {Quantum} {Driving} of a {Diamond} {Defect} {Ensemble}.
\newblock {\em Phys. Rev. Applied}, 15(5):054032, May 2021.
\newblock Publisher: American Physical Society.

\bibitem{57}
D.~Le~Sage, K.~Arai, D.~R. Glenn, S.~J. DeVience, L.~M. Pham, L.~Rahn-Lee,
  M.~D. Lukin, A.~Yacoby, A.~Komeili, and R.~L. Walsworth.
\newblock Optical magnetic imaging of living cells.
\newblock {\em Nature}, 496(7446):486--489, April 2013.
\newblock Number: 7446 Publisher: Nature Publishing Group.

\bibitem{75}
Benjamin~Rocco Moss.
\newblock {Nitrogen} {Vacancy} {Diamond} {Quantum} {Sensing} {Applied} to
  {Mapping} {Magnetic} {Fields} of {Bacteria}.
\newblock Master's thesis, University of Massachusetts Boston, United States --
  Massachusetts, 2021.
\newblock ISBN: 9798516097881.

\bibitem{1}
Eisuke Abe and Kento Sasaki.
\newblock Tutorial: {Magnetic} resonance with nitrogen-vacancy centers in
  diamond---microwave engineering, materials science, and magnetometry.
\newblock {\em Journal of Applied Physics}, 123(16):161101, March 2018.
\newblock Publisher: American Institute of Physics.

\bibitem{70}
Mason~C. Marshall, Reza Ebadi, Connor Hart, Matthew~J. Turner, Mark~J.H. Ku,
  David~F. Phillips, and Ronald~L. Walsworth.
\newblock High-{Precision} {Mapping} of {Diamond} {Crystal} {Strain} {Using}
  {Quantum} {Interferometry}.
\newblock {\em Phys. Rev. Applied}, 17(2):024041, February 2022.
\newblock Publisher: American Physical Society.

\bibitem{71}
Mason~C. Marshall, David~F. Phillips, Matthew~J. Turner, Mark J.~H. Ku, Tao
  Zhou, Nazar Delegan, F.~Joseph Heremans, Martin~V. Holt, and Ronald~L.
  Walsworth.
\newblock Scanning {X}-{Ray} {Diffraction} {Microscopy} for {Diamond} {Quantum}
  {Sensing}.
\newblock {\em Phys. Rev. Applied}, 16(5):054032, November 2021.
\newblock Publisher: American Physical Society.

\bibitem{97}
Jeong~Hyun Shim, Seong-Joo Lee, Santosh Ghimire, Ju~Il Hwang, Kwang-Geol Lee,
  Kiwoong Kim, Matthew~J. Turner, Connor~A. Hart, Ronald~L. Walsworth, and
  Sangwon Oh.
\newblock Multiplexed sensing of magnetic field and temperature in real time
  using a nitrogen vacancy spin ensemble in diamond.
\newblock {\em Phys. Rev. Applied}, 17(1):014009, January 2022.
\newblock Publisher: American Physical Society.

\bibitem{120}
Tatsuma Yamaguchi, Yuichiro Matsuzaki, Soya Saijo, Hideyuki Watanabe, Norikazu
  Mizuochi, and Junko Ishi-Hayase.
\newblock Control of all the transitions between ground state manifolds of
  nitrogen vacancy centers in diamonds by applying external magnetic driving
  fields.
\newblock {\em Jpn. J. Appl. Phys.}, 59(11):110907, November 2020.
\newblock Publisher: IOP Publishing.

\bibitem{123}
Chen Zhang, Heng Yuan, Ning Zhang, Lixia Xu, Jixing Zhang, Bo~Li, and Jiancheng
  Fang.
\newblock {Vector} magnetometer based on synchronous manipulation of
  nitrogen-vacancy centers in all crystal directions.
\newblock {\em J. Phys. D: Appl. Phys.}, 51(15):155102, March 2018.
\newblock Publisher: IOP Publishing.

\bibitem{86}
Andreas F.~L. Poulsen, Joshua~D. Clement, James~L. Webb, Rasmus~H. Jensen,
  Kirstine Berg-S{\o}rensen, Alexander Huck, and Ulrik~Lund Andersen.
\newblock Optimal control of a nitrogen-vacancy spin ensemble in diamond for
  sensing in the pulsed domain.
\newblock {\em arXiv:2101.10049 [physics, physics:quant-ph]}, January 2021.
\newblock arXiv: 2101.10049.

\bibitem{89}
Phila Rembold, Nimba Oshnik, Matthias~M. M{\"u}ller, Simone Montangero, Tommaso
  Calarco, and Elke Neu.
\newblock Introduction to quantum optimal control for quantum sensing with
  nitrogen-vacancy centers in diamond.
\newblock {\em AVS Quantum Sci.}, 2(2):024701, June 2020.
\newblock Publisher: American Vacuum Society.

\bibitem{92}
Romana Schirhagl, Kevin Chang, Michael Loretz, and Christian~L. Degen.
\newblock Nitrogen-vacancy centers in diamond: nanoscale sensors for physics
  and biology.
\newblock {\em Annu Rev Phys Chem}, 65:83--105, 2014.

\bibitem{30}
Koji Kobashi.
\newblock {\em Diamond {Films}: {Chemical} {Vapor} {Deposition} for {Oriented}
  and {Heteroepitaxial} {Growth}, Elsevier Science}.
\newblock 2010.

\bibitem{67}
N.~B. Manson, J.~P. Harrison, and M.~J. Sellars.
\newblock Nitrogen-vacancy center in diamond: {Model} of the electronic
  structure and associated dynamics.
\newblock {\em Phys. Rev. B}, 74(10):104303, September 2006.
\newblock Publisher: American Physical Society.

\bibitem{56}
J.~A. Larsson and P.~Delaney.
\newblock Electronic structure of the nitrogen-vacancy center in diamond from
  first-principles theory.
\newblock {\em Phys. Rev. B}, 77:165201, Apr 2008.

\bibitem{81}
Giulia Petrini, Ekaterina Moreva, Ettore Bernardi, Paolo Traina, Giulia
  Tomagra, Valentina Carabelli, Ivo~Pietro Degiovanni, and Marco Genovese.
\newblock Is a {Quantum} {Biosensing} {Revolution} {Approaching}?
  {Perspectives} in {NV}-{Assisted} {Current} and {Thermal} {Biosensing} in
  {Living} {Cells}.
\newblock {\em Advanced Quantum Technologies}, page 2000066, 2020.
\newblock \_eprint:
  https://onlinelibrary.wiley.com/doi/pdf/10.1002/qute.202000066.

\bibitem{26}
M.~W. Doherty, F.~Dolde, H.~Fedder, F.~Jelezko, J.~Wrachtrup, N.~B. Manson, and
  L.~C.~L. Hollenberg.
\newblock Theory of the ground-state spin of the nv${}^{\ensuremath{-}}$ center
  in diamond.
\newblock {\em Phys. Rev. B}, 85:205203, May 2012.

\bibitem{32}
{\'A}d{\'a}m Gali.
\newblock Ab initio theory of the nitrogen-vacancy center in diamond.
\newblock {\em Nanophotonics}, 8(11):1907--1943, November 2019.
\newblock Publisher: De Gruyter.

\bibitem{82}
L.~M. Pham, N.~Bar-Gill, D.~Le~Sage, C.~Belthangady, A.~Stacey, M.~Markham,
  D.~J. Twitchen, M.~D. Lukin, and R.~L. Walsworth.
\newblock Enhanced metrology using preferential orientation of nitrogen-vacancy
  centers in diamond.
\newblock {\em Phys. Rev. B}, 86(12):121202, September 2012.
\newblock Publisher: American Physical Society.

\bibitem{6}
Thiago P.~Mayer Alegre, Charles Santori, Gilberto Medeiros-Ribeiro, and
  Raymond~G. Beausoleil.
\newblock {Polarization}-selective excitation of nitrogen vacancy centers in
  diamond.
\newblock {\em Phys. Rev. B}, 76(16):165205, October 2007.
\newblock Publisher: American Physical Society.

\bibitem{60}
Till Lenz, Arne Wickenbrock, Fedor Jelezko, Gopalakrishnan Balasubramanian, and
  Dmitry Budker.
\newblock Magnetic sensing at zero field with a single nitrogen-vacancy center.
\newblock {\em Quantum Sci. Technol.}, 6(3):034006, June 2021.
\newblock Publisher: IOP Publishing.

\bibitem{77}
M.~Mr{\'o}zek, J.~Mlynarczyk, D.~S. Rudnicki, and W.~Gawlik.
\newblock Circularly polarized microwaves for magnetic resonance study in the
  {GHz} range: {Application} to nitrogen-vacancy in diamonds.
\newblock {\em Appl. Phys. Lett.}, 107(1):013505, July 2015.
\newblock Publisher: American Institute of Physics.

\bibitem{121}
Xiaoying Yang, Ning Zhang, Heng Yuan, Guodong Bian, Pengcheng Fan, and Mingxin
  Li.
\newblock Microstrip-line resonator with broadband, circularly polarized,
  uniform microwave field for nitrogen vacancy center ensembles in diamond.
\newblock {\em AIP Advances}, 9(7):075213, July 2019.

\bibitem{122}
Heng Yuan, Xiaoying Yang, Ning Zhang, Zhiqiang Han, Lixia Xu, Jixing Zhang,
  Guodong Bian, Pengcheng Fan, Mingxin Li, and Yuchen Liu.
\newblock {Frequency}-tunable and {Circularly} {Polarized} {Microwave}
  {Resonator} for {Magnetic} {Sensing} with {NV} {Ensembles} in {Diamond}.
\newblock {\em IEEE Sensors Journal}, pages 1--1, 2020.
\newblock Conference Name: IEEE Sensors Journal.

\bibitem{127}
Huijie Zheng, Arne Wickenbrock, Georgios Chatzidrosos, Lykourgos Bougas, Nathan
  Leefer, Samer Afach, Andrey Jarmola, Victor~M. Acosta, Jingyan Xu,
  Geoffrey~Z. Iwata, Till Lenz, Zhiyin Sun, Chen Zhang, Takeshi Ohshima,
  Hitoshi Sumiya, Kazuo Nakamura, Junichi Isoya, J{\"o}rg Wrachtrup, and Dmitry
  Budker.
\newblock Novel {Magnetic}-{Sensing} {Modalities} with {Nitrogen}-{Vacancy}
  {Centers} in {Diamond}.
\newblock {\em Engineering Applications of Diamond}, January 2021.
\newblock Publisher: IntechOpen.

\bibitem{128}
Huijie Zheng, Jingyan Xu, Geoffrey~Z. Iwata, Till Lenz, Julia Michl, Boris
  Yavkin, Kazuo Nakamura, Hitoshi Sumiya, Takeshi Ohshima, Junichi Isoya,
  J{\"o}rg Wrachtrup, Arne Wickenbrock, and Dmitry Budker.
\newblock Zero-{Field} {Magnetometry} {Based} on {Nitrogen}-{Vacancy}
  {Ensembles} in {Diamond}.
\newblock {\em Phys. Rev. Applied}, 11(6):064068, June 2019.
\newblock Publisher: American Physical Society.

\bibitem{43}
{Hincks, Ian}.
\newblock {\em Exploring {Practical} {Methodologies} for the {Characterization}
  and {Control} of {Small} {Quantum} {Systems}}.
\newblock {PhD} {Thesis}, UWSpace, 2018.

\bibitem{thesis}
{Liddy, Madelaine}.
\newblock {\em Optimal {Control} {Theory} {Techniques} for {Nitrogen} {Vacancy}
  {Ensembles}}.
\newblock {PhD} {Thesis}, UWSpace, 2022.

\bibitem{84}
F.~Poggiali, P.~Cappellaro, and N.~Fabbri.
\newblock Optimal control for one-qubit quantum sensing.
\newblock {\em Phys. Rev. X}, 8(2):021059, June 2018.
\newblock Publisher: American Physical Society.

\bibitem{REF1}
Kyryl Kobzar, Thomas~E Skinner, Navin Khaneja, Steffen~J Glaser, and Burkhard
  Luy.
\newblock Exploring the limits of broadband excitation and inversion pulses.
\newblock {\em J Magn Reson}, 170(2):236--243, Oct 2004.

\bibitem{REF2}
Soumyajit Mandal, Van D~M Koroleva, Troy~W Borneman, Yi-Qiao Song, and Martin~D
  H{\"u}rlimann.
\newblock Axis-matching excitation pulses for cpmg-like sequences in
  inhomogeneous fields.
\newblock {\em J Magn Reson}, 237:1--10, Dec 2013.

\bibitem{REF6}
Navin Khaneja, Timo~O. Reiss, Burkhard Luy, and Steffen~J. Glaser.
\newblock Optimal control of spin dynamics in the presence of relaxation.
\newblock {\em Journal of magnetic resonance}, 162 2:311--9, 2002.

\bibitem{REF3}
Troy~W Borneman, Martin~D H{\"u}rlimann, and David~G Cory.
\newblock Application of optimal control to cpmg refocusing pulse design.
\newblock {\em J Magn Reson}, 207(2):220--233, Dec 2010.

\bibitem{REF10}
Kyryl Kobzar, Sebastian Ehni, Thomas~E. Skinner, Steffen~J. Glaser, and
  Burkhard Luy.
\newblock Exploring the limits of broadband 90$\,^{\circ}$ and 180$\,^{\circ}$
  universal rotation pulses.
\newblock {\em Journal of Magnetic Resonance}, 225:142--160, 2012.

\bibitem{REF9}
Kyryl Kobzar, Thomas~E. Skinner, Navin Khaneja, Steffen~J. Glaser, and Burkhard
  Luy.
\newblock Exploring the limits of broadband excitation and inversion: {II}.
  {Rf}-power optimized pulses.
\newblock {\em Journal of Magnetic Resonance}, 194(1):58--66, 2008.

\bibitem{REF4}
Troy~W Borneman and David~G Cory.
\newblock Bandwidth-limited control and ringdown suppression in high-q
  resonators.
\newblock {\em J Magn Reson}, 225:120--129, Dec 2012.

\bibitem{REF5}
I.~N. Hincks, C.~E. Granade, T.~W. Borneman, and D.~G. Cory.
\newblock Controlling quantum devices with nonlinear hardware.
\newblock {\em Phys. Rev. Appl.}, 4:024012, Aug 2015.

\bibitem{REF7}
N.~Boulant, J.~Emerson, T.~F. Havel, D.~G. Cory, and S.~Furuta.
\newblock Incoherent noise and quantum information processing.
\newblock {\em J. Chem. Phys.}, 121(7):2955--2961, August 2004.
\newblock Publisher: American Institute of Physics.

\bibitem{REF8}
Marco~A. Pravia, Nicolas Boulant, Joseph Emerson, Amro Farid, Evan~M.
  Fortunato, Timothy~F. Havel, R.~Martinez, and David~G. Cory.
\newblock Robust control of quantum information.
\newblock {\em J. Chem. Phys.}, 119(19):9993--10001, November 2003.
\newblock Publisher: American Institute of Physics.

\bibitem{115}
Ralf Wunderlich, Robert Staacke, Wolfgang Knolle, Bernd Abel, J{\"u}rgen Haase,
  and Jan Meijer.
\newblock {Robust} nuclear hyperpolarization driven by strongly coupled
  nitrogen vacancy centers.
\newblock {\em Journal of Applied Physics}, 130(10):104301, September 2021.
\newblock Publisher: American Institute of Physics.

\bibitem{52}
Navin Khaneja, Timo Reiss, Cindie Kehlet, Thomas Schulte-Herbr{\"u}ggen, and
  Steffen~J. Glaser.
\newblock Optimal control of coupled spin dynamics: design of {NMR} pulse
  sequences by gradient ascent algorithms.
\newblock {\em Journal of Magnetic Resonance}, 172(2):296--305, February 2005.

\bibitem{114}
Christopher Wood, Ian Hincks, and Christopher Granade.
\newblock Quantumutils for mathematica, 2014.

\bibitem{Abragam}
A.~Abragam.
\newblock {\em The Principles of Nuclear Magnetism, Clarendon Press}.
\newblock 1961.

\bibitem{99}
Element Six.
\newblock {DNV}-{B1} 3.0x3.0mm, 0.5mm thick, June 2020.

\bibitem{100}
Element Six.
\newblock Unlocking next generation quantum technologies, June 2020.

\bibitem{5}
Victor~Marcel Acosta.
\newblock {\em Optical magnetometry with nitrogen-vacancy centers in diamond}.
\newblock PhD thesis, University of California, Berkley, Spring 2011.

\bibitem{73}
J~R Maze, A~Gali, E~Togan, Y~Chu, A~Trifonov, E~Kaxiras, and M~D Lukin.
\newblock Properties of nitrogen-vacancy centers in diamond: the group
  theoretic approach.
\newblock {\em New J. Phys.}, 13(2):025025, feb 2011.

\bibitem{110}
Amir Waxman.
\newblock {\em Sensitive {Magnetometry} {Based} on {NV} {Centers} in
  {Diamonds}}.
\newblock PhD thesis, University of the Negev, 2014.

\bibitem{98}
David~S Simon.
\newblock {\em Gaussian beams and lasers, Morgan and Claypool Publishers}.
\newblock 2053-2571. 2016.

\end{thebibliography}

\newpage

\section{Supplementary information}

\subsection{Mapping the Gaussian Beam in the Diamond Sample}

\begin{figure}[H] 
\centering
\includegraphics[width=0.7\textwidth]{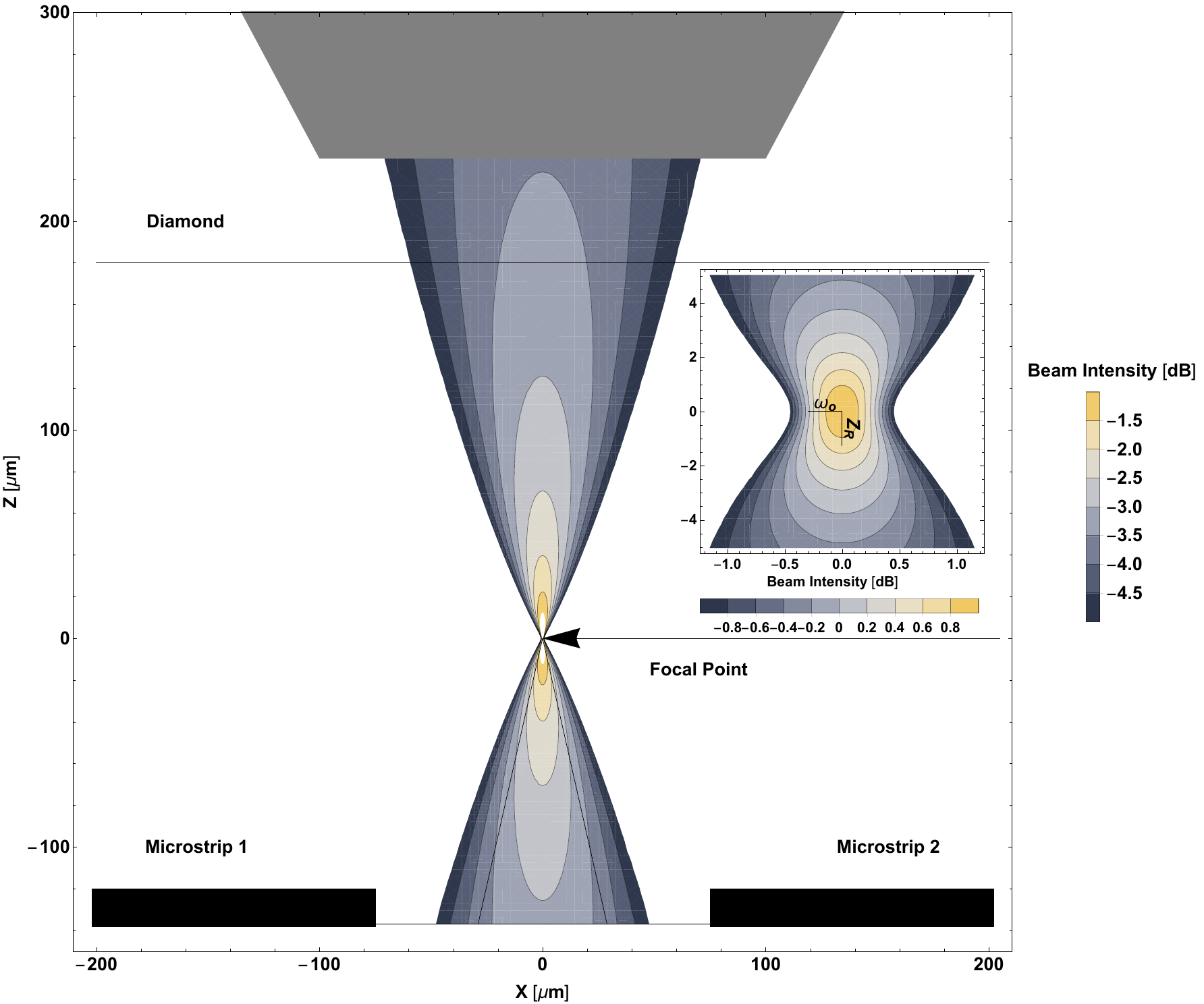}
\caption[Caption Summary]{The simulated image shows the Gaussian beam being focused into the diamond sample, indicating relative sizes of the diamond, PCB microstrips and NVs within the spot size of the objective. While there is swelling of the green light observed, the intensity of this quickly decays outside the focal volume. At the point the intensity of the green light hits the PCB surface, it is reduced by -3.5~dB to -5~dB from the maximum point at the focal point so reflections off the PCB may be neglected. Image Generated in Mathematica. \\
\textbf{Inset:} 
The intensity of a Gaussian Beam at the focal volume of the objective. The beam diameter of the 100x objective is twice the minimum beam waist, $\omega_o$, 0.59(4)~$mu$m. The depth of the field is twice the Rayleigh length $Z_R$, 2.51(4)~$\mu$m. At the Rayleigh length away, along the optical axis, the intensity of the beam is 50$\%$(0.69~dB) of the intensity at the maximum. At the edge of beam waist, the intensity is $\frac{1}{e^2}$ of the maximum at the center. The intensity of the beam drops off very quickly outside the depth of field and beam diameter, at a value of twice of each, it is already at 1$\%$(-1~dB) of the maximum intensity at the center of the focal volume. In this case only the NV centers within the focal volume are considered an effective part of the ensemble.}
\label{fig:1SI}
\end{figure}

In addition to the control field at the site of the NV focal volume shown in figure \ref{fig:2}, the Gaussian sites at the focal volume must be considered, and how the intensity of the beam changes as the beam diverges from the focal volume to the surface of the PCB board, as shown in figure \ref{fig:1SI}. There should be a high and uniform intensity at the focal volume, but the intensity should drop off quickly so that there is no background signal picked up from reflecting off the PCB board and microstrips. \\
    
The inset of figure \ref{fig:1SI} shows the intensity of the beam at the focal spot. With a 100x objective, this focal spot is quite small with beam diameter 0.59(4)~$\mu$m and beam depth of field 2.51(4)~$\mu$m, containing $\approx 16 000$ NV centers. As with all Gaussian beams, the intensity is halved at the Rayleigh range ($Z_R$) away from the center and $\frac{1}{e^2}$ at the beam radius ($\omega_o$) away from the center \cite{98}. Expanding beyond this, even at twice each of these values, the intensity of the beam has reduced by -1dB of the maximum intensity at the center of the focal volume. As the intensity dies off very quickly away from the center of the focal spot, only the NVs inside the focal spot are considered as the effective part of the ensemble. \\
    
The figure does show how the beam diverges well beyond the focal spot, however, due to the Gaussian nature of the beam, the intensity of the beam reduces quickly. Presented in a log scale, it is seen that at the point when the beam hits the PCB board, its intensity is already reduced by -3.5~dB to -5~dB. This further motivates using a thicker diamond ($>300~\mu$m) and objective with a shorter working distance ($160~\mu$m) to limit how close the focal volume may be to the PCB board, thus reducing the intensity sufficiently to make reflections off the PCB board negligible. To further reduce the background signal from reflections, a knotch and long pass filters are added inline with the emission path. 

\newpage
\subsection{Block Layout of the Optical Components}

\begin{figure}[H] 
\centering
\includegraphics[width=0.5\textwidth]{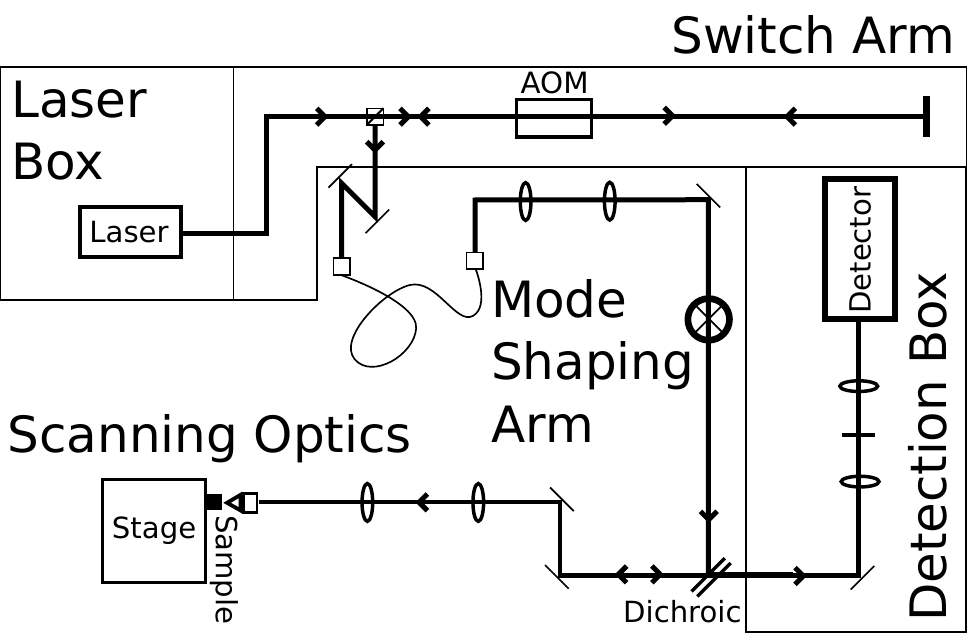}  \caption[Caption Summary]{The block layout of the optical setup. The laser box generates a continuous beam at 532nm. A double pass switch arm contains the AOM optical switch, used for its fast switching times and large contrast ratio between the ON and OFF intensity. The mode shaping arm corrects for beam shape aberrations as well as sets the desired optical polarization of the beam. The scanning optics focus the beam into the diamond sample, exciting the NVs with green light. The detector box collects the emitted red photons from the sample. Image generated in Inkscape.}
\label{fig:2SI}
\end{figure}

The block layout of the optical components in the NV ensemble setup is shown in figure \ref{fig:2SI}. This setup was designed to be easily adjusted to a single NV confocal setup should the experimental need arise. The block layout contains the laser box, switch arm, mode shaping arm, scanning optics and detection box as the main modules. \\

The laser box contains the source laser generating a beam at 532~nm, The switch arm contains an AOM as the fast optical switch in a double pass switch configuration to increase the ON:OFF contrast. The mode shaping arm corrects for spatial aberrations from the previous two modules and contains the half wave plate which is used to change the incoming optical polarization, used to equalize the fluorescence from the ensemble, shown in figure \ref{fig:3}. Following the mode shaping arm, the beam is directed to the objective with a dichroic mirror, chosen to reflect green light on the excitation path and transmit red on the emission path. The objective directs the green light to the sample, mounted atop a stage and collects the fluorescent red light emitted from the NVs, sending it to the detection box. The detection box contains a pinhole, a spatial filter for the emitted light followed by an Avalanche Photo Diode (APD), for photon detection. A more detailed account of the optical layout is available in \cite{thesis}.

\newpage
\subsection{The Block Layout of the RF Components}

\begin{figure}[H] 
\centering
\includegraphics[width=1\textwidth]{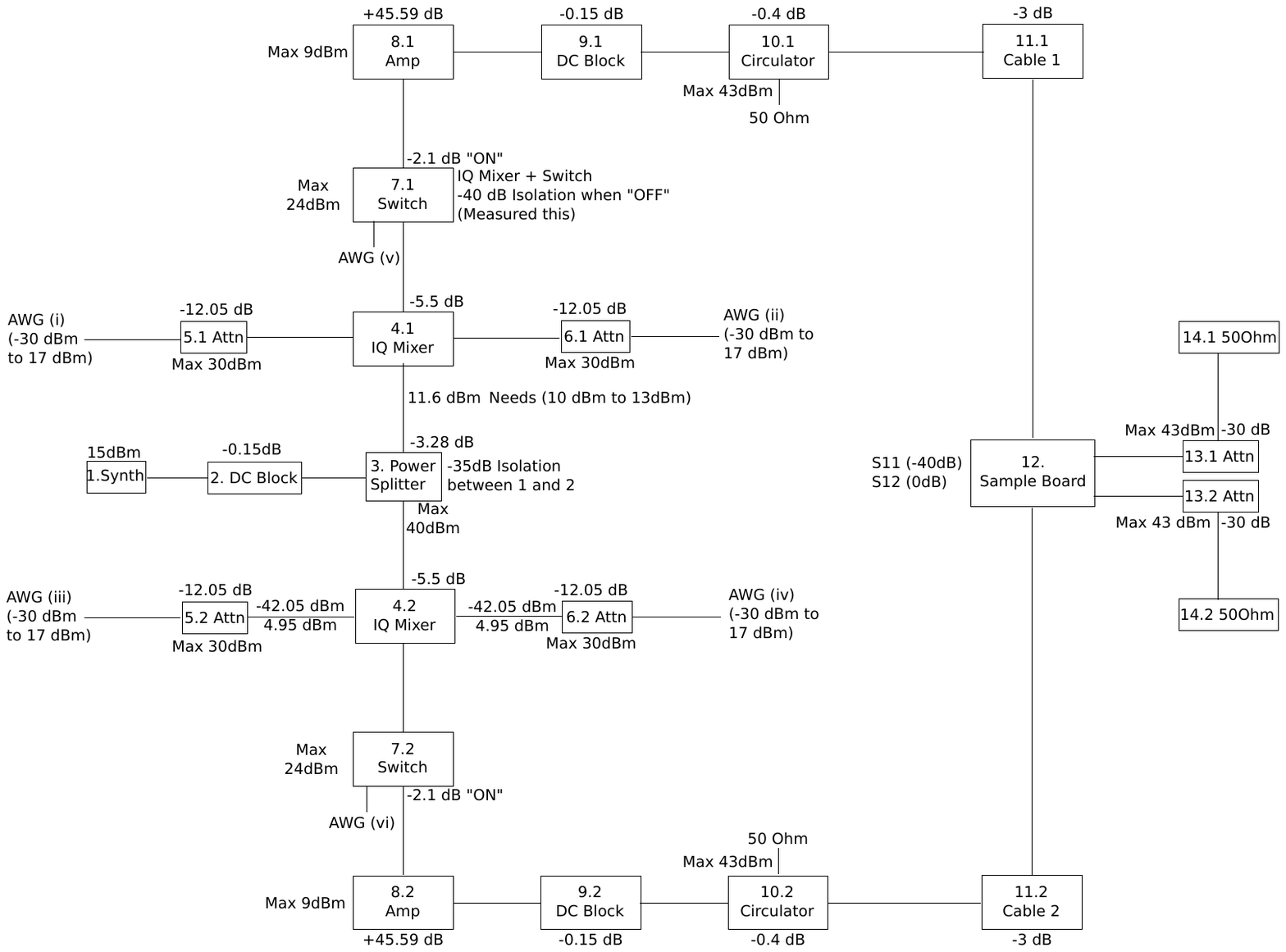}
\caption[Caption Summary]{The layout for a dual channel microwave control, capable of implementing circularly polarized microwaves. The frequency synthesizer provides the central frequency at $\omega_T$ (2.87~GHz), followed by a power splitter which divides the signal into two channels. The IQ mixer for each channel accepts the central frequency and combines with the amplitude and phase control envelopes from the AWG. The switch provides good isolation between the power between the ON and OFF state, while the amplifier boosts the amplitudes sent to the PCB board. The sample board contains two input ports for each channel, followed by two outputs and terminated into two attenuators. Image generated in Inkscape.}
\label{fig:3SI}
\end{figure}

Figure \ref{fig:3SI} shows the block layout for the RF components of the dual microwave system. The numbers shown in the layout are for indexing purposes to indicate the flow of information, and channel association (1 or 2). For each component, the figure indicates the output power [dBm] and any relevant power losses affecting the output power [dB]. Beginning with the frequency synthesizer, a central frequency at $2.87~$GHz is sent toward a power splitter, dividing the signal into two channels. This central frequency is then mixed with the envelopes sent by the four channel AWG into an IQ mixer. This signal which contains the control amplitude and phase at the central frequency passes through a switch, added for its increased isolation of the ON:OFF, and an amplifier to increase the strength of the signal before reaching the PCB sample board. The switch is placed before the amplifier in this case so as to not exceed the operating power threshold of the switch after the signal is amplified. With this configuration, there are two independent amplitude and phase controls available for experimentation to create the circularly polarized microwave control. For the best success of these dual-channel setups, identical components were used in order to minimize any differences between the two channels. A more detailed account of the complete RF signal and complete list of part numbers and manufacturer may be found in \cite{thesis}.
    
\begin{figure}[H] 
\centering
\includegraphics[width=0.3\textwidth]{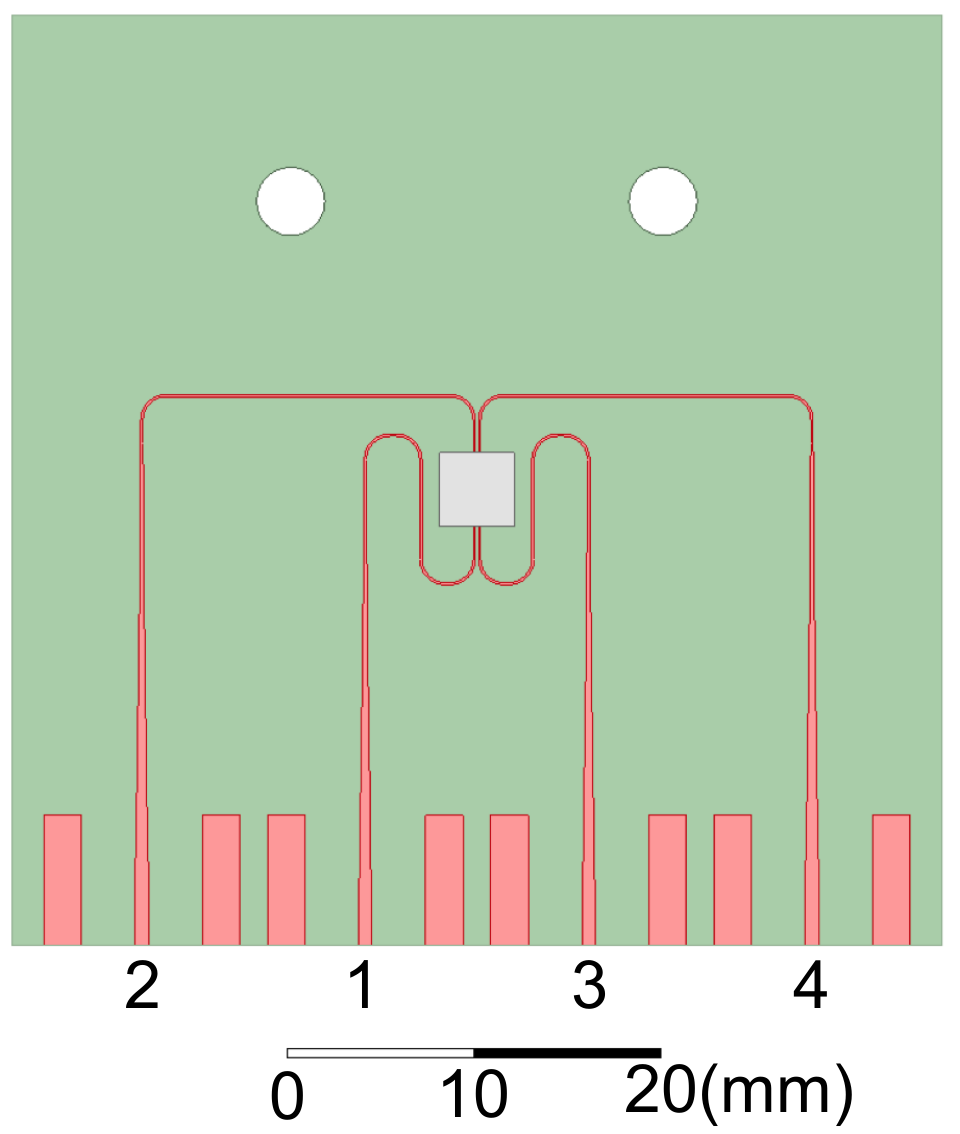}
\caption[Caption Summary]{The schematic of the dual-channel PCB board with a diamond mounted. The board was designed to have two inputs (ports 2 and 4) and two outputs (ports 1 and 3), and terminating with external $50~\Omega$ loads. The ports are labeled on the schematic above. The two microstrips are 150~$\mu$m apart, 127~$\mu$m wide, 17.5~$\mu$m thick and 7.5~mm long to accommodate diamond samples of larger size. The diamond sits on top of both microstrip, shown in the figure as a $4\times4$~mm diamond sample in the schematic. \textbf{Image Source:} Image courtesy of Hamid Mohebbi} 
\label{fig:4SI}
\end{figure}

Figure \ref{fig:4SI} shows the PCB board layout used in the experiments. There are two input (2 and 4) and two output ports (1 and 3) on the PCB, done to minimize the reflected power back to the amplifier and maximize the amount of power delivered into the diamond sample. A schematic of the diamond sample ($4\times4$~mm) is shown atop the two microstrips with the cross-section of the field diagram shown in figure \ref{fig:2}. The planar microstrips are 150~$\mu$m apart, 127~$\mu$m wide, 17.5~$\mu$m thick and 7.5~mm long to accommodate diamond samples of larger size. 

\newpage
\subsection{Spin-locking Simulation}
    
\begin{figure}[H]
\centering
\includegraphics[width=0.55\textwidth]{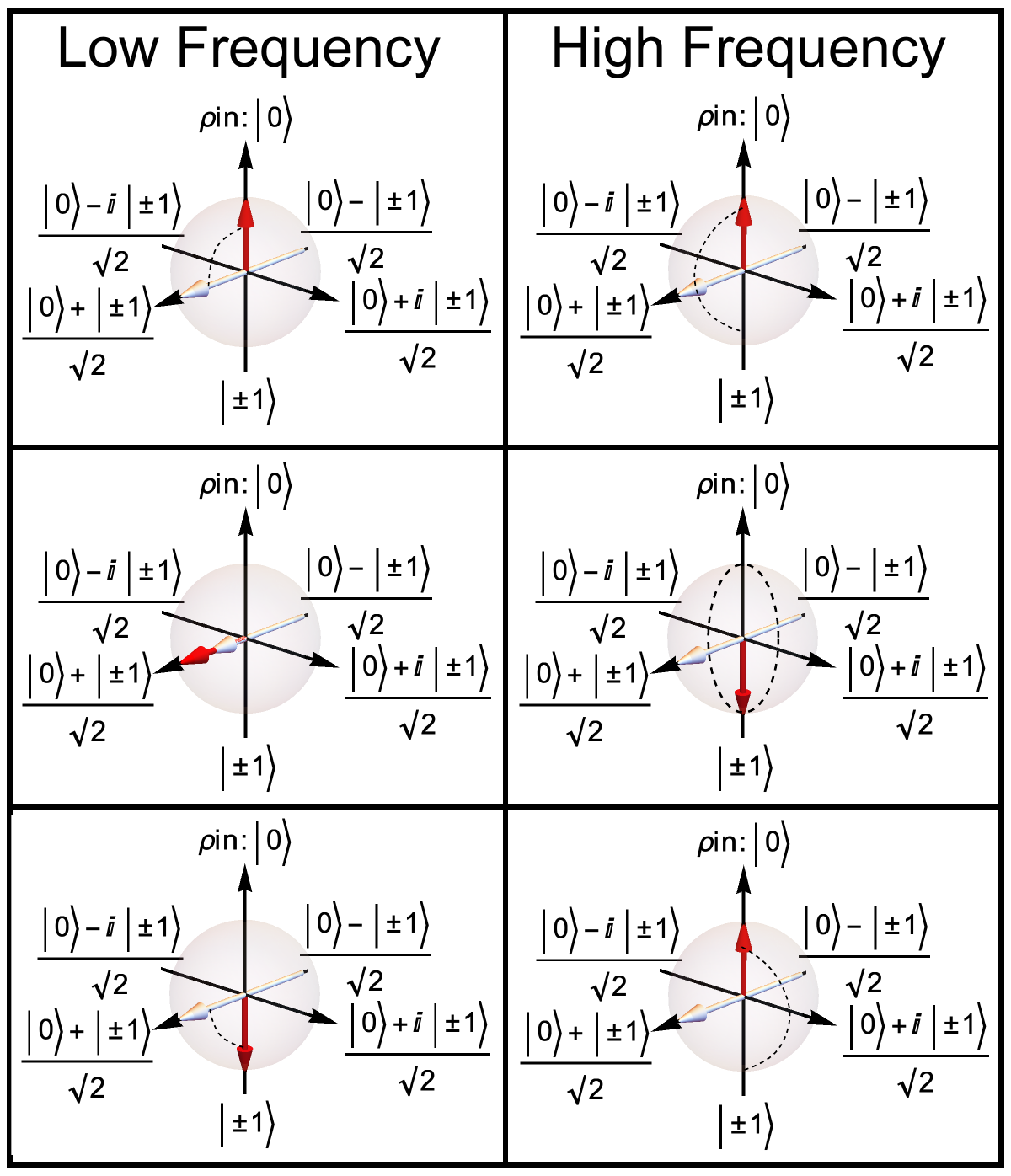}
\caption[Caption Summary]{The proof of concept for the spin-locking experiment is shown above in a Bloch sphere representation. Both the low and high-frequency sub-ensembles from the Rabi spectra begin in the $\ket{0}$ state \textbf{(red arrows)}. A square $\frac{\pi}{2}$ pulse targeted at the low-frequency pair is applied, rotating the low-frequency to the $\frac{\ket{0}+\ket{\pm1}}{\sqrt{2}}$ state (+``x" axis), and the high-frequency to the $\ket{\pm1}$ state. Applying a control Hamiltonian along the +``x" axis \textbf{(white arrow)}, will suppress the evolution of the low-frequency pair, while the high-frequency pair may evolve freely. After the evolution stage, another $\frac{\pi}{2}$ pulse is applied to rotate the low-frequency peak to the $\ket{\pm1}$ state while the high-frequency peak returns to the $\ket{0}$ state. The Spin-lock experiment measures the free evolution of the high-frequency while the low-frequency is suppressed. The same procedure may be applied to suppress the high-frequency peak while the low-frequency peak evolves. Image generated in Mathematica.}
\label{fig:5SI}
\end{figure}

The procedure of the spin-locking experiment is shown in figure \ref{fig:5SI}, corresponding to the data for the spin-locking experiment shown in figure \ref{fig:4}. Recalling that the spin-locking experiment is used to suppress the evolution of a chosen sub-ensemble and allow the evolution of another. \\

The figure shows how the low frequency sub-ensemble is suppressed while the high frequency evolves freely. For each stage, the red arrow is the starting state for each of the low and high frequency sub-ensemble, the dashed line indicated the action being performed in each stage while the white arrow is the direction of the control Hamiltonian. It shows the $\frac{\pi}{2}$ pulse, the length of which is gathered from the Rabi experiment, to suppress the low-frequency. This same pulse will rotate the high frequency by a full $\pi$. Then pulsing the control Hamiltonian on the relative ``+x" axis, suppresses the evolution of the low-frequency sub-ensemble while the high-frequency rotates freely. In the last stage, another $\frac{\pi}{2}$ is applied which rotates each sub-ensemble back to the ``z"-axis so a measurement may take place. The same procedure may be applied to suppress the high-frequency sub-ensemble and allow the low-frequency sub-ensemble to evolve. 

\newpage
\subsection{Sample OCT pulse} 

The GRAPE algorithm accepts the input control phase $\Delta\theta$, and measured experimental values $\Omega_{NV_{A/B}}$ and $\eta$, then optimizes over the free control parameters, $\Omega_{1/2}$ to arrive at the final pulse. The figure shows a sample OCT pulse, optimizing over the control amplitude. Each time step, the amplitude may range between $0 \rightarrow 1$ for optimization. The total length of the pulse ($10~\mu$s) is given by the discrete time step value (40ns), and number of steps (250). To mitigate hardware distortions, the pulse should be as smooth a possible, as is shown in the example. 

\newpage
\subsection{Bloch Trajectory of sub-ensemble pair B} 

\begin{figure}[H] 
\centering
\subfigure{\includegraphics[width=0.8\textwidth]{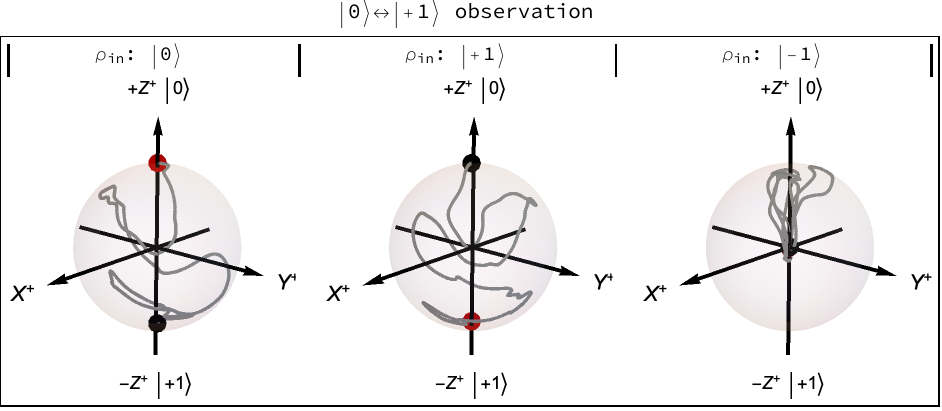}}
\subfigure{\includegraphics[width=0.8\textwidth]{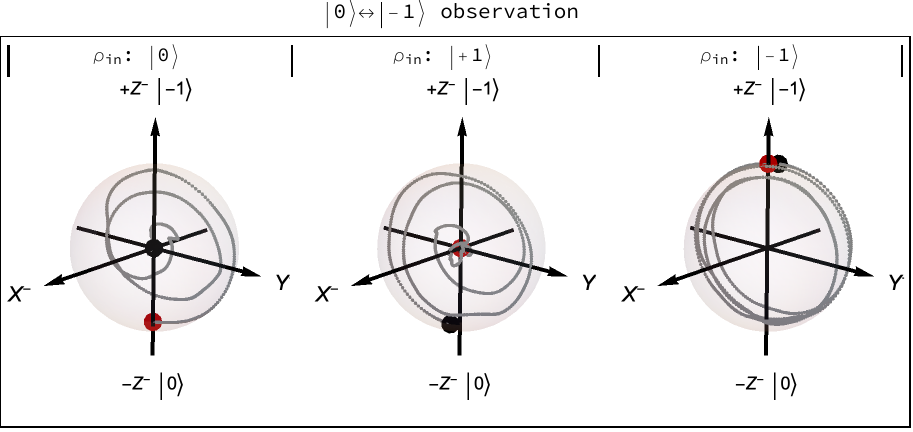}}
\caption[Caption Summary]{The Bloch Plot trajectories shown for the (1,1,1)$\&$(-1,1,-1) sub-ensemble corresponding to the same pulse optimized shown in figure \ref{fig:6} in the main text. As with the optimizing over the high-frequency sub-ensemble pair A shown in the main text, all the same successful transitions are observed for the selective $\pi_+$ pulse. An interesting observation is as this is the low-frequency pair, the trajectory does appear to be slower than the figure shown in the main text, to be expected as this is the low frequency sub-ensemble, so the amplitude envelope is less. Images generated in Mathematica.}
\label{fig:6SI}
\end{figure}

Figure \ref{fig:6SI} shows the Bloch plot trajectory for the NV sub-ensemble B, the lower frequency sub-ensemble, corresponding to the trajectory of sub-ensemble A shown in figure \ref{fig:6} in the main text. The same trends are observed as with the trajectory in sub-ensemble A, with a successful $\pi_+$ pulse demonstrated for varying input states. A notable difference is the trajectory does appear to be much slower, which is expected as the the NV envelope frequency ($\Omega_{NVB}$), guiding the pulse amplitude for this case is $1.03~$MHz compared to the $2.64$~MHz of the high-frequency sub-ensemble. 

\newpage
\subsection{The Complete Generalized Control \emph{a priori} Hamiltonian Model}

The phenomenological Hamiltonian provided a convenient experimental model that could be implemented for the proof of concept experiments. For a more complete theoretical Hamiltonian, an \emph{a priori} model may be used. It clearly separates the influence of the NV P.A.S. and microwave spatial components. This Hamiltonian is excellent to use in early simulations to observe the limitations on each of the NV P.A.S. and microwave spatial components and to model potential errors. \\

To create the \emph{a priori} Hamiltonian, first a frame rotation must take place so the Hamiltonian may be created with all the elements in the same frame. There are three important frames to consider when describing the \emph{a priori} Hamiltonian: the lab frame, the diamond crystal frame, and the NV P.A.S.  The ``z" axis of NV P.A.S. is defined as the vector between the Nitrogen and the Vacancy within the diamond crystal forms the (P.A.S.). The diamond crystal ``z" axis is indicated by the vector perpendicular to the diamonds' crystal frame. The lab frame is defined by the optical ``z" axis. \\

For the transformations, we may rotate either the microwave control field or the operators to the NV P.A.S. This frame is chosen as this contains the quantization access for the single NV or for a series of independent NVs in an ensemble. We choose to rotate the operators so the configuration of the microwave control field may be chosen at a later time. 
\begin{equation}
(R^T \cdot \vec{C}(t))\cdot\vec{S}=\vec{C}(t)(R\cdot\vec{S})
\label{eq:7}
\end{equation}
$\vec{C}(t)$ is the vector describing the direction of the microwave field applied. A field in the xz lab direction would be $\vec{C}(t)=\{C_x(t),0,C_z(t)\}$ and $\vec{S}$ is the vector representing the spin-1 operators $\vec{S}=\{S_x,S_y,S_z\}$. \\

A symbolic rotation matrix for factoring in the components of the NV P.A.S into the Hamiltonian may be used to start, the numeric values for the rotation matrix are displayed in figure \ref{fig:7SI} and may be subbed into the Hamiltonian when required for calculations. Applying the rotation matrix to the spin operators in their original frame expressed them in the rotated frame ($\vec{\tilde{S}}$) is shown below: 
\begin{equation}
R\vec{\tilde{S}}=\left(
\begin{array}{ccc}
R_{xx} & R_{xy} & R_{xz} \\
R_{yx} & R_{yy} & R_{yz} \\
R_{zx} & R_{zy} & R_{zz} \\
\end{array}
\right) \cdot (\tilde{S_{x}},\tilde{S_{y}},\tilde{S_{z}})^T  
\label{eq:4}
\end{equation}
Expressing the individual spin operators in the rotated frame as: 
\begin{equation}
\begin{split}
\tilde{S_{x}} & \rightarrow (R_{xx} S_{x} + R_{xy} S_{y} + R_{xz} S_{z}) \\
\tilde{S_{y}} & \rightarrow (R_{yx} S_{x} + R_{yy} S_{y} + R_{yz} S_{z}) \\ 
\tilde{S_{z}} & \rightarrow (R_{zx} S_{x} + R_{zy} S_{y} + R_{zz} S_{z}) \\ 
\end{split}
\label{eq:9}
\end{equation} 
Applying the rotation to the control field in the lab frame, a general control Hamiltonian with generic microwave field direction, in the NV frame is expressed as: 
\begin{equation}
\begin{split}
\CMcal{H}_{CTRL} & = \vec{C(t)}\cdot \vec{\tilde{S}} \\
\CMcal{H}_{CTRL} & = C_x(t) (R_{xx} S_{x} + R_{xy} S_{y} + R_{xz} S_{z}) \\
& + C_y(t) (R_{yx} S_{x} + R_{yy} S_{y} + R_{yz} S_{z}) \\
& + C_z(t)(R_{zx} S_{x} + R_{zy} S_{y} + R_{zz} S_{z}) 
\end{split}
\label{eq:10}
\end{equation}
At this point, the control Hamiltonian and NV ground state Hamiltonian are both expressed in the same NV frame. This general representation extends to any single NV. Now that they are both in the same frame, the effective Hamiltonian may be found for a single NV center.

\subsubsection{Finding the Effective Control Hamiltonian for a Single NV along any Principle Axis}

Now that the control Hamiltonian has been expressed in the frame of the NV P.A.S., the effective Hamiltonian may be found. The total Hamiltonian is the summation of the ZFS of the NV center and microwave control Hamiltonian. Recall that the rotation terms (xx,xy etc.) are unique for each NV orientation. 
\begin{equation}
\begin{split}
\CMcal{H}_{Tot} & = \Delta S_z^2 +  C_x(t) (R_{xx} S_{x} + R_{xy} S_{y} + R_{xz} S_{z}) \\
& + C_y(t) (R_{yx} S_{x} + R_{yy} S_{y} + R_{yz} S_{z}) + C_z(t)(R_{zx} S_{x} + R_{zy} S_{y} + R_{zz} S_{z}) 
\end{split}
\end{equation}
The following equation is used to find the effective Hamiltonian:  
\begin{equation}
\CMcal{H}_{Eff}=U^{\dagger}(t)(\CMcal{H}_{Tot}-\CMcal{H}_{Rot})U(t) 
\label{eq:Ham_eff_start}
\end{equation}
Where the rotation Hamiltonian $(\CMcal{H}_{Rot})$ is the P.A.S. of the NV center set at the transmitter frequency $(\omega_{T})$: 
\begin{equation}
\CMcal{H}_{Rot}= \omega_{T} S_z^2
\end{equation}
And the matrix exponential of which is:
\begin{equation}
U(t)=e^{-i\CMcal{H}_{Rot}t}=\mathds{1}-\left(1-e^{-i\omega_{T}t}\right)S_z^2
\end{equation}
The rotation Hamiltonian and its matrix exponential are substituted into equation \ref{eq:Ham_eff_start}:  
\begin{equation}
\begin{split}
\CMcal{H}_{Eff} & =\left(\mathds{1}-\left(1-e^{i\omega_{T}t}\right)S_z^2\right) \\
& \big((\Delta - \omega_{T})S_z^2 \\
& + C_x(t) (R_{xx}S_{x} + R_{xy}S_{y} + R_{xz}S_{z}) \\
& + C_y(t) (R_{yx}S_{x} + R_{yy}S_{y} + R_{yz}S_{z}) \\
& + C_z(t)(R_{zx}S_{x} + R_{zy}S_{y} + R_{zz}S_{z})\big) \\
& \left(\mathds{1}-\left(1-e^{-i\omega_{T}t}\right)S_z^2\right)
\end{split}
\label{eq:Ham_full}
\end{equation}
Equation \ref{eq:Ham_full} is expanded. To simplify, recall the following spin-1 operator relationships: 
\begin{equation}
\begin{split}
& S_z^{2n} = S_z^{2} \\
& S_z^{2n+1} = S_z \\
& S_z^2S_{(x/y)}S_z^2 = 0 
\end{split}
\end{equation}
Some trigonometric identities and Euler's formula may also be used to simplify: 
\begin{equation}
\begin{split}
(1- e^{i\omega_{T}t})(1- e^{-i\omega_{T}t}) & = 2-2\cos(\omega_{T}t) \\
e^{i \pm \omega_T t} & = \cos(\omega_T t) \pm i \sin(\omega_T t) \\
\end{split}
\end{equation}
Finally, using the commutators and anti-commutator relationships of the spin operators:
\begin{equation}
\begin{split}
\{S_{x/y/z},S_z^2\} & = S_{x/y} \\
\{(R_{xx}S_x+ R_{xy}S_y+ R_{xz}S_z),S_z^2\} & = (R_{xx}S_x+ R_{xy}S_y+ R_{xz}S_z) \\
\{(R_{yx}S_x+ R_{yy}S_y+ R_{yz}S_z),S_z^2\} & = (R_{yx}S_x+ R_{yy}S_y+ R_{yz}S_z) \\
\{(R_{zx}S_x+ R_{zy}S_y+ R_{zz}S_z),S_z^2\} & = (R_{zx}S_x+ R_{zy}S_y+ R_{zz}S_z) \\
[(R_{xx}S_x+ R_{xy}S_y+ R_{xz}S_z),S_z^2] & = R_{xx}[S_x,S_z^2]+R_{xy}[S_y,S_z^2] \\ 
[(R_{yx}S_x+ R_{yy}S_y+ R_{yz}S_z),S_z^2] & = R_{yx}[S_x,S_z^2]+R_{yy}[S_y,S_z^2] \\ 
[(R_{zx}S_x+ R_{zy}S_y+ R_{zz}S_z),S_z^2] & = R_{zx}[S_x,S_z^2]+R_{zy}[S_y,S_z^2] \\ 
\end{split}
\end{equation}
Further expanding and simplifying using the above relationships, the effective Hamiltonian for a general transmitter frequency $(\omega_T)$, microwave field direction $(C_x,C_y,C_z)$ and NV orientation (xx,xy, etc.) is found:
\begin{equation}
\begin{split}
\CMcal{H}_{Eff} & = (\Delta - \omega_T) S_z^2 \\
& + (R_{xz} C_x+R_{yx} C_y+R_{zz} C_z) S_z \\
& + (R_{xx} C_x + R_{yx} C_y + R_{zx} C_z) \left( \cos(\omega_T t)S_x -i\sin(\omega_T t)[S_x,S_z^2] \right)\\
& + (R_{xy} C_x + R_{yy} C_y + R_{zy} C_z ) \left(\cos(\omega_T t)S_y -i\sin(\omega_T t)[S_y,S_z^2] \right) \\
\end{split}
\label{eq:Ham_eff_Nosub}
\end{equation}
The microwave controls $(C_x,C_y,C_z)$ can now be expanded to give more context for the experiments conducted and abilities with each set of controls. From here, the ``twisted" operators will be substituted in for the commutation relationship, $S_{x/y}^{'}=i[S_{y/x},S_z^2]$. \\

Consider the case where the microwave field is controlled by two independent channels. Two independent channels were chosen as the combination of these will yield circularly polarized microwaves, allowing single selective transitions between the $\ket{0} \rightarrow \ket{+1}$ or $\ket{-1}$ states. \\ 

These two channels may both emit in the (x,y,z) directions. An example may be two microstrips which each have components in multiple directions. Under the assumption of two independent channels emitting along the (x,y,z) directions, $(C_x,C_y,C_z)$ may be expanded as: 
\begin{equation}
\begin{split}
C_x & = C_1(t)w_{x1} + C_2(t)w_{x2} \\
C_y & = C_1(t)w_{y1} + C_2(t)w_{y2} \\
C_z & = C_1(t)w_{z1} + C_2(t)w_{z2} \\
\end{split}
\label{eq:C_expanded}
\end{equation}
The quantities $w_{x1/2}$ describes the x-component of the field for channels 1 and 2, respectively. This is similar for $w_{y1/2}$ and $w_{z1/2}$. These fields are defined by the configuration of the control field, but may be expressed in the general case above. \\

$C_{1/2}(t)$ represent the time dependent controls shaped by the AWG, IQ mixer and set at the transmitter frequency. In an ideal case, $C_{1/2}(t)$ are shown in equation \ref{eq:17}. Here the AWG controls are shown in Cartesian coordinates $I_{1/2}$ and $Q_{1/2}$, but this may also be represented by polar coordinates with amplitude and phase control; $\Omega_{1/2}$ and $\theta_{1/2}$, where $I_{1/2}=\Omega_{1/2}\cos(\theta_{1/2})$ and $Q_{1/2}=\Omega_{1/2}\sin(\theta_{1/2})$.
\begin{equation}
C_{1/2}(t)=I_{1/2}(t)\cos(\omega_Tt)+Q_{1/2}(t)\sin(\omega_Tt) 
\label{eq:17}
\end{equation}
To further simply, the squares of the cosine and sine functions may be expanded:
\begin{equation}
\begin{split}
\cos^2(\omega_Tt) & = \frac{1}{2}(1+\cos(2 \omega_Tt)) \\
\cos(\omega_Tt)\sin(\omega_Tt) & =\frac{1}{2}\sin(2 \omega_Tt) \\
\sin^2(\omega_Tt) & = \frac{1}{2}(1-\cos(2 \omega_Tt))
\end{split}
\end{equation}
Expanding the controls for each channel, $(C_x,C_y,C_z)$, into the Hamiltonian in equation \ref{eq:Ham_eff_Nosub}, and using the trigonometric identities listed above yields the generalized effective Hamiltonian. In this instance, the values for I and Q are being analyzed at each discrete time step, so the time-dependency has been dropped. \\
\begin{equation}
\begin{split}
\CMcal{H}_{eff} & =(\Delta - \omega_T) S_z^2  \\
& + \left(\cos(\omega_T t)(I_1w_{A1} + I_2w_{A2}) + \sin(\omega_Tt)(Q_1w_{A1} + Q_2w_{A2})\right)S_z \\
& + \left(\frac{1}{2}(1+\cos(2\omega_Tt))(I_1w_{B1}+I_2w_{B2}) + \frac{1}{2}\sin(2\omega_Tt)(Q_1w_{B1}+Q_2w_{B2})\right)S_x \\
& - \left( \frac{1}{2}\sin(2\omega_Tt)(I_1w_{B1}+I_2w_{B2}) +\frac{1}{2}(1-\cos(2\omega_Tt))(Q_1w_{B1}+Q_2w_{B2}) \right)S_y^{'}\\
& + \left(\frac{1}{2}(1+\cos(2\omega_Tt))(I_1w_{C1}+I_2w_{C2}) + \frac{1}{2}\sin(2\omega_Tt)(Q_1w_{C1}+Q_2w_{C2})\right)S_y \\
& - \left( \frac{1}{2}\sin(2\omega_Tt)(I_1w_{C1}+I_2w_{C2}) +\frac{1}{2}(1-\cos(2\omega_Tt))(Q_1w_{C1}+Q_2w_{C2}) \right)S_x^{'} \\
\end{split}
\end{equation}

$w_{A1/2}$, $w_{B1/2}$ and $w_{C1/2}$ represent the microwave field components and NV rotational terms, gathered to reduce the complexity of the Hamiltonian.  
\begin{equation}
\begin{split}
w_{A1/2} & = xz w_{x1/2} + yz w_{y1/2} + zz w_{z1/2} \\
w_{B1/2} & = xx w_{x1/2} + yx w_{y1/2} + zx w_{z1/2} \\
w_{C1/2} & = xy w_{x1/2} + yy w_{y1/2} + zy w_{z1/2} \\
\end{split}
\end{equation}
As the experiments are all performed in the absence of a magnetic field, the center transmitter frequency $(\omega_T)$, is set to the ZFS $(\Delta)$. All resulting terms then proportional to $(2\Delta)$ may be dropped. The terms proportional to $S_z$ also become negligible, as they would induce a Zeeman splitting, proportional to the strength of the microwave controls, but oscillating at 2.87~GHz, and so would be averaged out compared to the other more slowly varying terms. \\

This yields the time-independent effective Hamiltonian for two independently controlled channels for a generalized field in the $(x,y,z)$ directions. The Hamiltonian is presented with letter terms $(A,B,C,D)$ for simplicity.
\begin{equation}
\begin{split}
\CMcal{H}_{eff} & = A S_x + B S_y^{'} + C S_y + D S_x^{'} \\
& A = \frac{1}{2}(I_1w_{B1}+I_2w_{B2})\\
& B = - \frac{1}{2}(Q_1w_{B1}+Q_2w_{B2}) \\
& C =  \frac{1}{2}(I_1w_{C1}+I_2w_{C2}) \\
& D = - \frac{1}{2}(Q_1w_{C1}+Q_2w_{C2}) 
\label{eq:ham_1}
\end{split}
\end{equation}
Finally, expanding the values for the microwave components $(w_{B1})$ etc. yields the full Hamiltonian. 
\begin{equation}
\begin{split}
\CMcal{H}_{eff} & = A S_x + B S_y^{'} + C S_y + D S_x^{'} \\
& A = \frac{1}{2}(R_{xx} (I_1w_{x1} +I_2w_{x2})+ R_{yx} (I_1w_{y1} +I_2w_{y2}) + R_{zx} (I_1w_{z1}+I_2 w_{z2})) \\
& B = - \frac{1}{2}(R_{xx} (Q_1w_{x1} +Q_2w_{x2})+ R_{yx} (Q_1w_{y1} +Q_2w_{y2}) + R_{zx} (Q_1w_{z1}+Q_2 w_{z2})) \\
& C =  \frac{1}{2}(R_{xy} (I_1w_{x1} +I_2w_{x2})+ R_{yy} (I_1w_{y1} +I_2w_{y2}) + R_{zy} (I_1w_{z1}+I_2 w_{z2})) \\
& D = - \frac{1}{2}(R_{xy} (Q_1w_{x1} +Q_2w_{x2})+ R_{yy} (Q_1w_{y1} +Q_2w_{y2}) + R_{zy} (Q_1w_{z1}+Q_2 w_{z2})) 
\end{split}
\label{eq:22}
\end{equation}

Equation \ref{eq:ham_1} shows the basic structure with the letter format. Each pre-factor $(A,B,C,D)$ is dependent on the AWG envelope of control $(I_{1/2},Q_{1/2})$, direction and strength of the microwave field at the site of the NVs, $(w_{x1/2},w_{y1/2},w_{z1/2})$ for each independent channel, and last by the NV rotation term, ($R_{xx}$,$R_{xy}$,$R_{xz}$ etc.). \\

Although this Hamiltonian describes a single NV, from here it is clear to see the difficulty in controlling ensembles of NV centers as the projection of the effective Hamiltonian is scaled by the orientation of the NV center, and if the volume of NVs is large, the microwave field strength and direction of each also scales over the volume. \\

In the following section, the Hamiltonian will be manipulated, without loss of generality to show how to account for the projection of the Hamiltonian into each NV orientation in a more mathematically convenient way for achieving single transitions.

\subsubsection{Expressing the Control Hamiltonian with Pseudo Spin-1/2 Operators}

To ease the calculations in solving for selective ground state transitions in the Hamiltonian, the spin-1 operators may be written as a sum of pseudo spin-$\frac{1}{2}$ operators. These are of course not true spin-$\frac{1}{2}$ operators, as the ground state $\ket{\pm 1}$ states share a space with the same $\ket{0}$ state. Expressing the spin-1 operators in this form, allows for the selective transitions between $\ket{0} \rightarrow \ket{+1}/\ket{-1}$ in the absence of a magnetic field, to be found more easily. \\

The pseudo spin-$\frac{1}{2}$ operators are labelled as $S_x^{\pm}$ and $S_y^{\pm}$ for the $\ket{+1}$ and $\ket{-1}$ pseudo sub-spaces, respectively. The operators are shown below, recalling $S_{x/y}^{'}=i[S_{y/x},S_z^2]$:
\begin{equation}
\begin{split}
S_x^+ & = \frac{1}{\sqrt{2}}(S_x + i[S_y,S_z^2]) \\
S_y^+ & = \frac{1}{\sqrt{2}}(S_y - i[S_x,S_z^2]) \\
S_x^- & = \frac{1}{\sqrt{2}}(S_x - i[S_y,S_z^2]) \\
S_y^- & = \frac{1}{\sqrt{2}}(S_y + i[S_x,S_z^2]) \\
S_z^+ & = \frac{1}{2} S_z + \mathds{1}_3 - \frac{3}{2} S_z^2 \\
S_z^- & = \frac{1}{2} S_z - \mathds{1}_3 + \frac{3}{2} S_z^2 \\
\end{split}
\end{equation}
In matrix form, the resemblance to the Pauli operators can be seen, as is the intention. 
\begin{equation}
\begin{split}
S_x^+ & = 
\left(
\begin{array}{ccc}
0 & 0 & 0 \\
0 & 0 &  1 \\
0 & 1 & 0 
\end{array}
\right) ;
S_y^+= 
\left(
\begin{array}{ccc}
0 & 0 & 0 \\
0 & 0 &  -i \\
0 & i & 0 
\end{array}
\right) ; 
S_z^+ = 
\left(
\begin{array}{ccc}
0 & 0 & 0 \\
0 & 1 &  0 \\
0 & 0 & -1 
\end{array}
\right) \\
S_x^- & = 
\left(
\begin{array}{ccc}
0 & 1 & 0 \\
1 & 0 & 0 \\
0 & 0 & 0 
\end{array}
\right) ;
S_y^-= 
\left(
\begin{array}{ccc}
0 & -i & 0 \\
i & 0 & 0 \\
0 & 0 & 0 
\end{array}
\right) ;
S_z^- = \left(
\begin{array}{ccc}
1 & 0 & 0 \\
0 & -1 &  0 \\
0 & 0 & 0 
\end{array}
\right) ;
\label{eq:pseudo_spin_1_2_operators}
\end{split}
\end{equation}
The Hamiltonian in equation \ref{eq:22} has already been grouped in accordance to the pseudo spin-$\frac{1}{2}$ operators, so substituting the pseudo spin-$\frac{1}{2}$ operators, and collecting to isolate for these are trivial. Re-arranging the pseudo spin-$\frac{1}{2}$ operators in the original spin-1 form, the following is substituted into the Hamiltonian: \\
\begin{equation}
\begin{split}
S_x & = \frac{1}{\sqrt{2}}(S_x^+ + S_x^-) \\
S_y & = \frac{1}{\sqrt{2}}(S_y^+ + S_y^-) \\
i[S_x,S_z^2] & = \frac{1}{\sqrt{2}}(S_y^- - S_y^+) \\
i[S_y,S_z^2] & = \frac{1}{\sqrt{2}}(S_x^+ - S_x^-) 
\end{split}
\end{equation} 
Now re-arranging for the pseudo spin-$\frac{1}{2}$ operators, without loss of generality, the Hamiltonian is written with the pseudo spin-$\frac{1}{2}$ operators. New pre-factors $(\tilde{A},\tilde{B},\tilde{C},\tilde{D})$ are used to distinguish between the pre-factors for the Hamiltonian written in equation \ref{eq:22}:
\begin{equation}
\begin{split}
\CMcal{H} & =\tilde{A}S_x^+ + \tilde{B}S_y^+ + \tilde{C}S_x^- + \tilde{D}S_y^- \\
\tilde{A} & = \frac{1}{2\sqrt{2}}(I_1w_{B1}+I_2w_{B2}-Q_1w_{C1}-Q_2w_{C2}) \\
\tilde{B} & = \frac{1}{2\sqrt{2}}(I_1w_{C1}+I_2w_{C2}+Q_1w_{B1}+Q_2w_{B2}) \\
\tilde{C} & = \frac{1}{2\sqrt{2}}(I_1w_{B1}+I_2w_{B2}+Q_1w_{C1}+Q_2w_{C2}) \\
\tilde{D} & = \frac{1}{2\sqrt{2}}(I_1w_{C1}+I_2w_{C2}-Q_1w_{B1}-Q_2w_{B2}) 
\end{split}
\end{equation}
The full Hamiltonian written by expanding the terms $w_{B1/2}$ and $w_{C1/2}$, is shown below:
\begin{equation}
\begin{split}
\CMcal{H} = &\tilde{A}S_x^+ + \tilde{B}S_y^+ + \tilde{C}S_x^- + \tilde{D}S_y^- \\
\tilde{A}  = & \frac{1}{2\sqrt{2}}(R_{xx}(I_1w_{x1}+I_2w_{x2})+R_{yx}(I_1w_{y1}+I_2w_{y2})+R_{zx}(I_1w_{z1}+I_2w_{z2}) \dots \\
& -R_{xy}(Q_1w_{x1}+Q_2w_{x2})-R_{yy}(Q_1w_{y1}+Q_2w_{y2})-R_{zy}(Q_1w_{z1}+Q_2w_{z2})) \\
\tilde{B} = & \frac{1}{2\sqrt{2}}(R_{xy}(I_1w_{x1}+I_2w_{x2})+R_{yy}(I_1w_{y1}+I_2w_{y2})+R_{zy}(I_1w_{z1}+I_2w_{z2}) \dots \\
& +R_{xx}(Q_1w_{x1}+Q_2w_{x2})+R_{yx}(Q_1w_{y1}+Q_2w_{y2})+R_{zx}(Q_1w_{z1}+Q_2w_{z2})) \\
\tilde{C} = & \frac{1}{2\sqrt{2}}(R_{xx}(I_1w_{x1}+I_2w_{x2})+R_{yx}(I_1w_{y1}+I_2w_{y2})+R_{zx}(I_1w_{z1}+I_2w_{z2}) \dots \\
& +R_{xy}(Q_1w_{x1}+Q_2w_{x2})+R_{yy}(Q_1w_{y1}+Q_2w_{y2})+R_{zy}(Q_1w_{z1}+Q_2w_{z2})) \\
\tilde{D} = & \frac{1}{2\sqrt{2}}(R_{xy}(I_1w_{x1}+I_2w_{x2})+R_{yy}(I_1w_{y1}+I_2w_{y2})+R_{zy}(I_1w_{z1}+I_2w_{z2}) \dots \\
& -R_{xx}(Q_1w_{x1}+Q_2w_{x2})-R_{yx}(Q_1w_{y1}+Q_2w_{y2})-R_{zx}(Q_1w_{z1}+Q_2w_{z2})) \\
\label{eq:27}
\end{split}
\end{equation}

\newpage
\subsubsection{Substituting in the values for the NV P.A.S.}

\begin{figure}[H]
\centering
 \subfigure{\textbf{(a)}\includegraphics[width=0.75\textwidth]{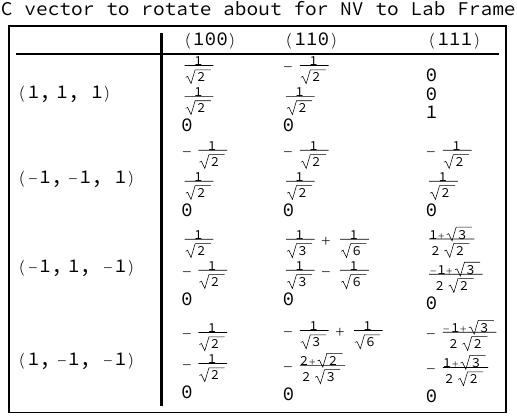}}
 \subfigure{\textbf{(b)}\includegraphics[width=0.75\textwidth]{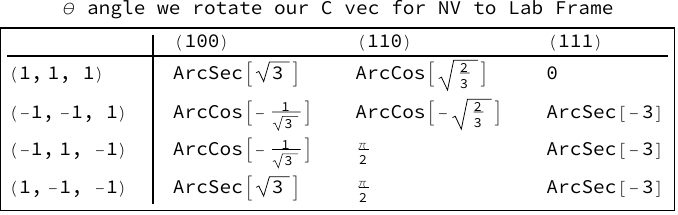}}
 \subfigure{\textbf{(c)}\includegraphics[width=1\textwidth]{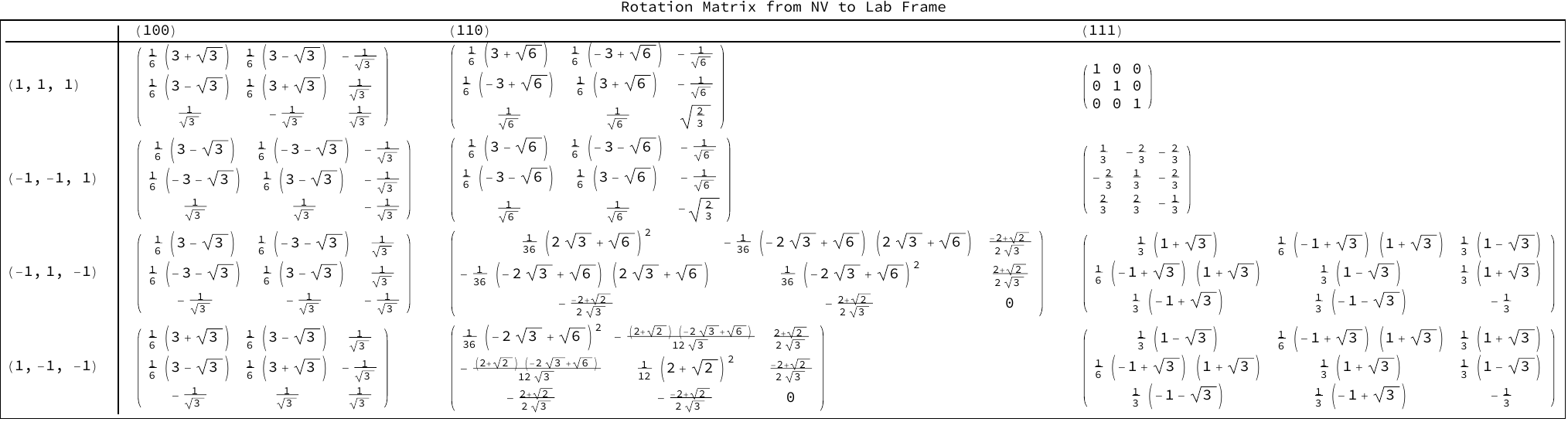}}
   \caption[NV to Lab Frame Rotation]{\textbf{(a)} and \textbf{(b)} The rotation vector and angle to rotate the NV (P.A.S.) to the lab frame for all four orientations in each of the (100), (110) and (111) diamond. \textbf{(c)} Rotation matrix from the lab frame to the NV frame for all NVs in three crystals. NVs are labelled (1,1,1), (-1,-1,1), (-1,1,-1) and (1,-1,-1). Diamonds are labelled (100), (110) and (111). Images generated on Mathematica.}
   \label{fig:7SI}
\end{figure}

The values for the generic rotation matrix shown in equation \ref{eq:4}, are shown in figure \ref{fig:7SI}, generated by performing the frame transformation procedure below. 
\begin{enumerate}
\item Normalize the starting vector being rotated (labelled $\vec{a}$)
\item Find the cross product between the starting vector ($\vec{a}$) and the desired vector ($\vec{b}$) to yield the rotation axis ($\vec{c}$)
\item Find the dot product between $\vec{a}$ and $\vec{b}$ to yield the rotation angle ($\theta_{(ab)}$)
\item Create a rotation matrix, which rotates $\vec{a}$ about $\vec{c}$ by $\theta_{(ab)}$ to arrive at $\vec{b}$
\end{enumerate}

Using this technique, the rotation matrices for the diamond crystal to lab and NV to crystal frame may be found. These matrices are listed below for each NV orientation for the (100), (110) and (111) crystals. These three crystals are shown as they are the common crystal orientations used for NV experiments, and expressed here to show how this can be applied to any single crystal orientation. These values may then be substituted into the control Hamiltonian in equation \ref{eq:22} and \ref{eq:27} for a specific NV orientation and crystal. 

These rotation values may be substituted along with the microwave spatial components corresponding to the field diagram in figure \ref{fig:2} to arrive at the final \emph{a priori} Hamiltonian. 

\end{document}